%% file: mainJSTA-finalarxiv.tex
\documentclass[11pt, english]{paper}
\usepackage{latexsym}
\usepackage{amsmath}
\usepackage{amsfonts}
\usepackage{amssymb}
\usepackage{psfrag}
\usepackage{graphicx}
\usepackage{natbib}
\usepackage[normalem]{ulem} 
\usepackage[colorlinks=true,citecolor=blue]{hyperref}
\usepackage[margin=1in]{geometry}
\usepackage{setspace} 

\newtheorem{theorem}{Theorem}
\newtheorem{assumption}{Assumption}
\newtheorem{cor}{Corollary}
\newtheorem{lem}{Lemma}

\newtheorem{prop}{Proposition}
\newtheorem{definition}{Definition}
\newtheorem{remark}{Remark}
\newtheorem{ex}{Example}

\def\qed{\hspace{3mm} \rule{2mm}{2mm}}

\renewcommand{\eqref}[1]{(\ref{#1})}

\newcommand{\bfu}{\mathbf {u}}
\newcommand{\bfv}{\mathbf {v}}

\newcommand{\bfy}{\mathbf {y}}

\newcommand{\bfK}{\mathbf {K}}

\newcommand{\bT}{\mathbf {T}}

\newcommand{\bZ}{\mathbf {Z}}
\newcommand{\cA}{\mathcal {A}}

\newcommand{\cD}{\mathcal {D}}

\newcommand{\cH}{\mathcal {H}}

\newcommand{\cK}{\mathcal {K}}
\newcommand{\cL}{\mathcal {L}}
\newcommand{\cO}{\mathcal {O}}

\newcommand{\cS}{\mathcal {S}}
\newcommand{\cV}{\mathcal {V}}

\newcommand{\cW}{\mathcal {W}}

\newcommand{\pr}{\stackrel {pr}{\longrightarrow}}

\def\bve{{\boldsymbol\ve}}

\def\bbeta{{\boldsymbol\eta}}
\def\bgamma{{\boldsymbol\gamma}}
\def\bGamma{{\boldsymbol\Gamma}}

\def\bmu{{\boldsymbol\mu}}
\def\bnu{{\boldsymbol\nu}}

\def\btheta{{\boldsymbol\theta}}
\def\bTheta{{\boldsymbol\Theta}}

\newcommand{\cov}{\mbox{cov}}

\def\cov{\mbox{\textnormal{cov}}}

\def\disp{\displaystyle}

\newcommand{\real}{\ensuremath{\mathbb{R}}}

\newcommand{\dC}{\ensuremath{\mathbb{C}}}

\newcommand{\dF}{\ensuremath{\mathbb{F}}}

\newcommand{\dI}{\ensuremath{\mathbb{I}}}

\newcommand{\dK}{\ensuremath{\mathbb{K}}}

\newcommand{\dR}{\ensuremath{\mathbb{R}}}

\newcommand{\dZ}{\ensuremath{\mathbb{Z}}}
\def\ve{\varepsilon}

\def\bx{\mathbf{x}}
\def\bX{\mathbf{X}}

\def\bu{\mathbf{u}}
\def\bU{\mathbf{U}}
\def\by{\mathbf{y}}

\def\bv{\mathbf{v}}
\def\bV{\mathbf{V}}

\def\be{\mathbf{e}}
\def\bF{\mathbf{F}}
\newcommand{\cF}{\ensuremath{\mathcal{F}}}
\def\sn{ n^{\frac{1}{2}} }
\def\sni{ n^{-\frac{1}{2}} }
\def\bve{{\boldsymbol\ve}}

\def\bGamma{{\boldsymbol\Gamma}}
\def\bmu{{\boldsymbol\mu}}

\def\btheta{{\boldsymbol\theta}}

\def\bTheta{{\boldsymbol\Theta}}

\def\bfl{{\boldsymbol\ell}}
\def\setd{\{1,\ldots,d\}}
\def\setn{\{1,\ldots,n\}}

\newcommand{\I}{\ensuremath{\boldsymbol{1}}}

\def\unif{{\rm U}(0,1)}

\title{On testing for independence between generalized error models of several time series}

\author{Kilani Ghoudi$^{\;1}$ \and Bouchra R. Nasri$^{\;2}$ \and Bruno N. R\'emillard$^{\;1,\;3}$\thanks{Corresponding author: bruno.remillard@hec.ca}}
\institution{
$^1$Department of Statistics and Business Analytics, United Arab Emirates University, \\P.O.Box 15551, Al Ain, UAE \\ 
$^2$D\'epartement de m\'edecine sociale et pr\'eventive, \'Ecole de sant\'e publique,\\ Universit\'e de Montr\'eal, C.P. 6128, succursale Centre-ville
Montr\'eal (Qu\'ebec)  H3C 3J7 \\ 
$^3$Department of Decision Sciences, HEC Montr\'eal, \\3000 chemin de la C\^ote Sainte-Catherine,
Montr\'eal (Qu\'ebec), Canada H3T 2A7
}

\begin{document}

\maketitle

\begin{abstract}
We define generalized innovations associated with generalized error models  having  arbitrary distributions, that is, distributions that can be mixtures of continuous and discrete distributions. These models include stochastic volatility models and regime-switching models with possibly zero-inflated regimes. The main novelty of this article is to be able to test for conditional independence between several time series with arbitrary distributions, extending previous results of Duchesne, Ghoudi \& R\'emillard (2012), obtained only for stochastic volatility models. We define families of empirical processes constructed from lagged generalized errors, and we show that their joint asymptotic distributions are Gaussian and independent of the estimated parameters of the individual time series. M\"obius transformations of the empirical processes are used to obtain tractable covariances.
Several tests statistics are then proposed, based on Cram\'er–von Mises type statistics and dependence measures, as well as graphical methods to visualize the dependence. In addition,  numerical experiments are performed to assess the power of the proposed tests. Finally, to show the usefulness of our methodologies, examples of applications for financial data and crime data are given to cover both discrete and continuous cases. All developed methodologies are implemented in the CRAN package IndGenErrors \citep{Ghoudi/Nasri/Remillard/Duchesne:2025}.
\end{abstract}


\section{Introduction}

Testing conditional independence between time series is crucial for understanding and modeling their relationships. For example, if the time series $X_{1t}$ and $X_{2t}$ are conditionally independent with respect to the past, denoted by
\begin{equation}\label{eq:cond_ind}
 P\left(X_{1t}\le x_1, X_{2t}\le x_2 |\cF_{t-1}\right) = P\left(X_{1t}\le x_1 |\cF_{t-1}\right) P\left(X_{2t}\le x_2 |\cF_{t-1}\right) = G_{1t}(x_1)G_{2t}(x_2),
\end{equation}
then, this is equivalent to saying that $X_{2t}$ can be ignored for predicting $X_{1t}$, or is not a leading indicator \citep{Haugh:1976}, meaning that
\begin{equation}\label{eq:cond_ind2}
P(X_{1t}\le x_1|\cF_{t-1},X_{2t}=x_2) = P(X_{1t}\le x_1|\cF_{t-1}).
\end{equation}
In fact, if $G_{2t}$ is discontinuous at $x_2$, then \eqref{eq:cond_ind} implies that $P(X_{1t}\le x_1,X_{2t}=x_2|\cF_{t-1}) = G_{1t}(x_1)P(X_{2t}=x2)$, showing that \eqref{eq:cond_ind2} holds, and if the density of $G_{2t}$ is continuous and positive at $x_2$, then \eqref{eq:cond_ind} also implies that \eqref{eq:cond_ind2} holds true. One can even remove the continuity assumption by using properties of conditional expectations.
According to \cite{Haugh:1976}, finding out if one time series can act or not as a leading indicator for the other is one of the main motivations for checking the independence between time series. It is also the first step in trying to model multivariate time series, since having independence between ``innovations'' of time series simplifies the underlying model.
In addition, if the conditional distribution functions $G_{1t}$ and $G_{2t}$ are continuous, then \eqref{eq:cond_ind} is equivalent to having independence between the innovations $U_{1t}=G_{1t}(X_{1t})$ and $U_{2t}=G_{2t}(X_{2t})$. Note that these innovations are independent and uniformly distributed in $(0,1)$, as noted in \cite{Bai:2003}. The independence test between innovations can serve as a diagnostic tool to discover if there is remaining dependence, and if so, add a covariate to reduce dependence, as exemplified in Section \ref{ssec:nasdaq}.

Several tests of dependence between innovations have been proposed in the literature that depend mainly on the properties of the data and the characteristics of the proposed models. Basically, there are three methods for testing independence: using cross-correlations or other dependence measures, using empirical processes, and using tests on coefficients of multivariate models.

For example, \cite{Haugh:1976} proposed a test that is based on cross-correlations at different lags between residuals of two ARMA time series models. A similar approach was later proposed by \cite{Hong:1996} using kernel methods to be able to consider all possible lags; see also \cite{Duchesne/Roy:2003} for a robust approach based on Hong's idea. However, it is known that
cross-correlation-based methods can detect dependence for Gaussian models but they can fail to detect dependence for non-Gaussian models; see, e.g.,  the results for the so-called Tent Map in Table 1 of \cite{Duchesne/Ghoudi/Remillard:2012}.

To overcome the problem of estimating dependence using cross-correlations, \cite{Kim/Lee:2005} proposed the use of statistics based on empirical processes constructed from residuals of time series models, by comparing the joint empirical distribution function with the product of the empirical margins at different lags. Their work was carried out in the context of infinite autoregressive bivariate models. One can also think of copula-based methods as a natural way to characterize dependence. This approach was proposed in \cite{Genest/Remillard:2004} for
independent and identically (i.i.d.) observations and was later adapted to time series in \cite{Genest/Ghoudi/Remillard:2007} and \cite{Duchesne/Ghoudi/Remillard:2012}. In the latter case, stochastic volatility models were fitted to data, taking into account exogenous variables, and empirical processes of lagged residuals were used for testing independence between lagged innovations. However, stochastic volatility models do not cover the whole range of attractive time series models. For example, one could be interested in fitting regime-switching models \citep{Remillard:2013}.

Finally, a third way of testing independence between multivariate time series is to fit a multivariate model such as the Dynamic Conditional Correlation GARCH model \citep{Engle:2002}, or Vector autoregressive models (VAR), and test the null hypothesis that some coefficients are zero. In general, these coefficients are chosen so that the estimation of individual time series gives the same results as the joint estimation when the series are conditionally independent. For other models, such as VAR, the null hypothesis of independence between innovations is often equivalent to choosing diagonal matrices of parameters. For example, if $\bX_t = \bmu + A\bX_{t-1}+\bve_t$, with $\bve_t\stackrel{i.i.d.}{\sim} N_d(0,\Omega)$,  the null hypothesis of independence would be that $\Omega$ is diagonal. Again, note that it does not mean that the series are independent; only that their innovations $\ve_{1t},\ldots,\ve_{dt}$ are independent, or equivalently that the series are conditionally independent given the past.

The previous articles all dealt with the continuous case, where the distribution of each time series is assumed to be continuous. However, in many interesting cases, time series can be discrete or a mixture of discrete and continuous distributions, as in the case of zero-inflated models. To model this kind of dependence, there are a few models, such as the INGARCH model introduced in \cite{Ferland/Latour/Oraichi:2006}; see also \cite{Davis/Holan/Lund/Ravishanker:2016} and references therein. For example, based on a bivariate Poisson model introduced by \cite{Lakshminarayana/Pandit/Rao:1999}, \cite{Cui/Zhu:2018} studied independence between two count data series, where the joint distribution is given by
$\disp
P(X_1 = x_1, X_2 = x_2) = f_{\lambda_1}(x_1)f_{\lambda_2}(x_2)\left\{ 1+\delta\left(e^{-x_1}-e^{-cx_1}\right) \left(e^{-x_2}-e^{-cx_2}\right)\right\}$,
where $f_\lambda$ is the density of a Poisson distribution with parameter $\lambda>0$, $\disp |\delta| \le \frac{1}{\left(1-e^{-cx_1}\right)\left(1-e^{-cx_2}\right)}$, with $c=1-e^{-1}$. In this case, the independence is equivalent to $\delta=0$.

However, in general, it is difficult to produce multivariate time series models, apart from these particular models, especially if one of the series does not have a continuous distribution. Recently, in the continuous case, \cite{Nasri/Remillard:2019} introduced copula-based models for multivariate time series, where (continuous) univariate time series were combined by using copulas between ``innovations'' or ``generalized errors''. Recall that for any $d$-dimensional random vector $X$  with joint distribution function $H$ and margins $F_1,\ldots,F_d$, there exists a joint distribution function $C$ with uniform margins, so that for all $x_1,\ldots,x_d \in \dR$,
\begin{equation}\label{eq:sklar0}
H(x_1,\ldots,x_d)=P(X_1\le x_1,\ldots, X_d\le x_d) = C\left\{F_1(x_1),\ldots,F_d(x_d)\right\}.
\end{equation}
Here, $C$ is called a copula, and if the margins $F_1,\ldots,F_d$ are continuous, then $C$ is the joint distribution of the random vector $\bU=(U_1,\ldots,U_d)$, where
$U_j=F_j(X_j)$, $j\in\setd$. However, if one of the margins is not continuous, an infinite number of copulas satisfy \eqref{eq:sklar0}. Fortunately, whatever the margins, if $\bU$ has distribution $C$, denoted $\bU \sim C$, for a copula $C$ satisfying \eqref{eq:sklar0}, then $\left(F_1^{-1}(U_1),\ldots, F_d^{-1}(U_d)\right) \sim H$. Thus, a copula $C$ and margins $F_1,\ldots,F_d$ can be combined to recover any joint distribution function. This idea can be adapted for building multivariate time series models with arbitrary margins. This methodology is not new for non-continuous distributions. In fact, it was used to simulate dependent Markov chains in \cite[Section 8.2]{Remillard:2013}, as well as for simulating multivariate data with arbitrary distributions in \cite{Genest/Neslehova/Remillard/Murphy:2019}. For inferential purposes, this technique was also applied recently in \cite{Debaly/Truquet:2023} for modeling dependence between time series, with either continuous or discrete margins, while here we can use time-varying copulas and arbitrary distributions as well.

In what follows, using recent results on the construction of the multilinear copula \citep{Genest/Neslehova/Remillard:2017}, we present  a way for constructing ``innovations''  for multivariate time series with arbitrary distributions, and we propose tests of independence between lagged series of these innovations. We adopt a two-stage approach in the sense that individual time series are modeled and time-varying copulas are used to introduce additional dependence between series, on top of the dependence due to exogenous variables. This particular construction of multivariate copula-based models is described in Section \ref{sec:multi}. Next, in Section \ref{sec:emp}, the main theoretical results on empirical processes for testing independence are presented. In particular, we show that the methodology proposed in \cite{Duchesne/Ghoudi/Remillard:2012}, obtained in the special case of stochastic volatility models, can be applied to the new models. We also extend the results of \cite{Kheifets/Velasco:2017} for univariate discrete data to our multivariate setting. In Section \ref{sec:stats}, the main test statistics are presented, and numerical experiments are performed in Section \ref{sec:sim} to assess the finite sample performance (level and power) of the proposed statistics for different models. Finally, examples of applications are given in Section \ref{sec:applications}.

\section{Multivariate time series models}\label{sec:multi}

 The main idea for modeling  univariate times series is similar to the one used to define generalized error models \citep{Du:2016} in the continuous case. In fact, for a time series $X_t$,  a generalized error model assumes the existence of a series of random mappings $H_t$ so that $  \ve_{t} = H_{t}(X_t)$  are i.i.d. with common continuous margin $F$, and the $\ve_t$ can be called ``innovations''. Here, $H_t$ is a $\cF_{t-1}$-measurable function, where  $\cF_{t-1}$ contains information about the past values $X_{t-1},\ldots$, and potentially information about exogenous variables.  Note that generalized error models contain stochastic volatility models as special cases, where $H_t(y) = \frac{y-\mu_t}{\sigma_t}$, $\mu_t$ and $\sigma_t^2$ are, respectively, the conditional mean and the conditional variance given $\cF_{t-1}$.   Other interesting examples include
  the setting proposed by \cite{Bai:2003}, where $G_t(y)=P(X_t\le y|\cF_{t-1})$ is a measurable mapping not depending on $X_t$, so $\epsilon_t=G_t(X_t)$ is well defined and has a uniform distribution.
  In fact, for $u\in (0,1)$, $P\{G_t(X_t)\le u|\cF_{t-1}\} = P\left\{ X_t \le G_t^{-1}(u) |\cF_{t-1}\right\} = G_t\circ G_t^{-1}(u)=u$ if $G_t$ is continuous.
  In this case, $F=D$, where $D$ is the uniform distribution function. These models are particularly well suited for   regime-switching models where there is no natural innovations, and include the previous models by taking $G_t = F\circ H_t$.

  \begin{remark}\label{rem:gof}
  For specifications tests related to such models, see, e.g., \cite{Bai:2003} and \cite{Kheifets:2015} for continuous cases, and \cite{Kheifets/Velasco:2013,Kheifets/Velasco:2017} for discrete cases.
  \end{remark}

  In what follows, being interested in arbitrary distributions, e.g., a Poisson distribution, we need another approach, since the $G_t(X_t)$ are not necessarily independent or uniformly distributed. To solve this problem, we shall use a randomization technique proposed by \cite{Brockwell:2007}. More precisely, let $X_t$ be a given time series and let $\cV_t$ be a sequence of i.i.d. uniform variables independent of $X_t$. Then, define the sequence of i.i.d. uniform variables $U_t$ by setting
\begin{equation}
\label{eq:extge}
  U_{t} = \cH_{t}(X_t,\cV_t)=G_t(X_t-)+\cV_t \Delta G_t(X_t),
\end{equation}
where $G_t(y)=P(X_t\le y|\cF_{t-1})$, $G_t(y-)=P(X_t < y|\cF_{t-1})$, and $\Delta G_t(y)=G_t(y)-G_t(y-)=P(X_t=y|\cF_{t-1})$.
It is well known \citep{Ferguson:1967,Brockwell:2007, Neslehova:2007} that the $U_t$s are uniformly distributed conditionally to the past, meaning that $P(U_t\le u|\cF_{t-1})=u$. To prove that they are independent, note that using the properties of conditional expectations,
\begin{eqnarray*}
P(U_1 \le u_1, \ldots, U_n \le u_n) &=&  E\left[ E\left[\left. \prod_{j=1}^{n} \I\{U_j\le u_j\} \right| \cF_{n-1})\right]\right] \\
&=&
E\left[ \prod_{j=1}^{n-1} \I\{U_j\le u_j\} P(U_n \le u_n |\cF_{n-1})\right] = u_n P( U_1\le u_1,\ldots, U_{n-1}\le u_{n-1}).
\end{eqnarray*}
Repeating the same argument $n-1$ times, we end up with
$P(U_1 \le u_1, \ldots, U_n \le u_n) =\disp \prod_{j=1}^n u_j$, completing the proof.
Note that if $G_t$ is continuous, then $U_t=G_t(X_t)$, that is, there is no randomization effect. Hence, our proposed model is an extension of Bai's setting. To simplify the exposition, somewhat imprecisely, we will call the $U_t$s  generalized errors as well.

\begin{ex}\label{ex:bernoulli}
For example, let $X_t$ is a Markov chain on $\{0,1\}$ with transition matrix $Q = \left(\begin{array}{cc} p & 1-p \\ 1-p & p\end{array}\right)$, $p\in (0,1)$, and initial value $X_0=1$. In this case, $ G_t(0) =P(X_t=0|\cF_{t-1}) = p \I(X_{t-1}=0)+ (1-p)\I(X_{t-1}=1)$,  $\Delta G_t(0)=G_t(0)=G_t(1-)$, $G_t(0-)=0$, and $\Delta G_t(1) =P(X_t=1|\cF_{t-1}) = (1-p) \I(X_{t-1}=0)+ p \I(X_{t-1}=1)$. As a result, using \eqref{eq:extge}, one gets
\begin{eqnarray*}
U_t &=& \cH_t(X_t,\cV_t) = p \cV_t \I(X_{t-1}=0,X_t=0) +(p +(1-p)\cV_t)\I(X_{t-1}=0,X_t=1) \\
&& \qquad  + (1-p) \cV_t \I(X_{t-1}=1,X_t=0) +  (1-p +p \cV_t)\I(X_{t-1}=1,X_t=1).
\end{eqnarray*}
It follows that for any $u\in [0,1]$, $P(U_t\le u|\cF_{t-1}) = u$, proving that the ``generalized innovations'' $U_t$ are i.i.d. and uniformly distributed.
\end{ex}
Finally, to define the dependence between time series, let $\tilde \bU_t=(\tilde U_{1t},\ldots, \tilde U_{dt})$ be a sequence of random vectors with uniform $(0,1)$ margins, so that $P(\tilde \bU_t\le \bu|\cF_{t-1})=C_t(\bu)$, where $\cF_t$ is a filtration which might contain exogenous random variables, and with respect to which $\tilde\bU_t$ is measurable. Examples of time-varying copulas $C_t$ include parametric copula families with time-varying parameters \citep{Nasri/Remillard/Bouezmarni:2019} and regime-switching copulas \citep{Fink/Klimova/Czado/Stoeber:2017,Nasri/Remillard/Thioub:2020}.  Then, for conditional cdf $G_{jt}$, $j\in\setd$, set $\tilde X_{jt}=G_{jt}^{-1}(\tilde U_{jt})$, $t\in\setn$.  It then follows that we have the so-called conditional Sklar's equation \citep{Patton:2004}, i.e.,
$\disp 
    P\left(\tilde X_{1t}\le y_1,\ldots, \tilde X_{dt}\le y_d|\cF_{t-1}\right) =
    C_t\left\{G_{1t}(y_1),\ldots,G_{dt}(y_d)\right\}$.
In particular, if $\bU_t=(U_{1t},\ldots,U_{dt})$, where $U_{jt}=\cH_{jt}(X_{jt},\cV_{jt})$, and $
\bnu_t=(\cV_{1t},\ldots,\cV_{dt})\sim \Pi$, with $\Pi$ the independence copula, then $P(\bU_t\le \bu|\cF_{t-1}) = C_t^\maltese(\bu)$, where
$C_t^\maltese$ is called the multilinear copula \citep{Genest/Neslehova/Remillard:2017,Nasri/Remillard:2024c}, and $C_t^\maltese$ also satisfies the conditional Sklar's equation, 
\begin{equation}\label{eq:sklarmaltese}
    P(X_{1t}\le y_1,\ldots, X_{dt}\le y_d|\cF_{t-1}) = C_t^\maltese\left\{G_{1t}(y_1),\ldots,G_{dt}(y_d)\right\}.
\end{equation}
Recall that if all margins $G_{jt}$ are continuous, then $C_t$ is unique and $C_t=C_t^\maltese$.
Observe that the time series  $X_{1t},\ldots, X_{dt}$ are conditionally independent iff $C_t^\maltese =\Pi$ for all $t\in\setn$. This is our null hypothesis.

\begin{remark}\label{rem:null}
Under this null  hypothesis, by using properties of conditional expectations, it follows that for any $\bfl\in\dZ^d$, and conditionally to the past,
$\bU_{t+\bfl} = (U_{1,t+\ell_1},\ldots,U_{d,t+\ell_d})\sim \Pi$; here, for $t+\bfl$, the past is $\cF_{t-1+\min\{\ell_1,\ldots,\ell_d\}}$.
To show that this is true, assume without loss of generality that $0=\ell_1 \le \ell_2 \le \cdots \le \ell_d$. Then, we have to show that
$\disp
P(U_{1,t}=U_{1,t+\ell_1}\le u_1, \ldots, U_{d,t+\ell_d}\le u_d|\cF_{t-1})= \prod_{j=1}^d u_j$.
If $\ell_d=0$, there is nothing to prove. So suppose that $k = \max\{\ell_1,\ldots,\ell_d \} >0$ and set $m =\min\{1\le j\le d: \ell_j=k\}$.
 Using the null hypothesis that for any $s\ge 1$, $P(U_{1,s}\le u_1, \ldots,U_{d,s}\le u_d|\cF_{s-1})= \prod_{j=1}^d u_j$, we have
\begin{eqnarray*}
 P(U_{1,t}&=&U_{1,t+\ell_1}\le u_1, \ldots,U_{d,t+\ell_d}\le u_d|\cF_{t-1})\\
 &=&
E\left[
\I\{U_{1,t}\le u_1\} \cdots \I\{U_{m-1,t+\ell_{m-1}}\le u_{m-1}\}  P\left( U_{m,t+\ell_m}\le u_m,\ldots, U_{d,t+\ell_d}\le u_d|\cF_{t+k-1}\right)
|\cF_{t-1}\right]\\
&=& E\left[
\I\{U_{1,t}\le u_1\} \cdots \I\{U_{m-1,t+\ell_{m-1}}\le u_{m-1}\}  P\left( U_{m,t+k}\le u_m,\ldots, U_{d,t+k}\le u_d|\cF_{t+k-1}\right)
|\cF_{t-1}\right]\\
&=& \left\{\prod_{j=m}^d u_j \right\}
P\left(
U_{1,t}\le u_1, \ldots, U_{m-1,t+\ell_{m-1}}\le u_{m-1} |\cF_{t-1}\right).
\end{eqnarray*}
To complete the proof, we just have to repeat the previous arguments whenever $\max\{\ell_1,\ldots,\ell_{m-1} \} >0$.
\end{remark}
Testing the conditional between lagged innovations might be interesting in applications to detect if two series are conditionally dependent for some lag. For example, the effect of a decrease of oil price can affect the price of gas at the pump after some days. Therefore, from now on, based on Remark \ref{rem:null}, we consider the following null hypothesis of independence
\begin{equation*}\label{eq:H0}
    H_0: C_t^\maltese \sim \Pi, \quad \text{ for any } t\ge 0,
\end{equation*}
while the alternative hypothesis is
\begin{equation*}\label{eq:H1}
    H_1: C_t^\maltese \sim \Pi, \quad \text{ for some } t\ge 0.
\end{equation*}
In a forthcoming paper, we will present tests of specifications for the general multilinear conditional copulas $C_t^\maltese$, but here, we restrict our attention to the specific case of a constant copula, i.e., $C_t^\maltese=C$, for all $t>0$ and for some copula $C$.
\section{Empirical processes for testing independence}\label{sec:emp}

In what follows, we consider parametric univariate models of the form $H_{\btheta,jt}$ with continuous unknown margin $F_j$, or the extended generalized error models of the form $\cH_{\btheta,jt}$, with continuous margin $F_j=D$. For simplicity, set $\cH_{\btheta,jt}(y_j,v_j)=H_{\btheta,jt}(y_j)$ and define the generalized errors $\ve_{jt}=\cH_{\btheta_0,jt}(X_{jt},\cV_{jt})$.  Since exogenous variables may be common to several series, all parameters are grouped in a meta-parameter $\btheta_0$, and it is assumed that $\btheta_n$ is a consistent estimator of $\btheta_0$ such that $\bTheta_n=\sn(\btheta_n-\btheta_0)$ converges in law to $\bTheta$, denoted $\bTheta_n \rightsquigarrow \bTheta$.
For simplicity, set $G_{jt} = G_{\btheta_0,jt}$.
Next,  define the pseudo-observations $e_{n,jt}= \cH_{\btheta_n,jt}(X_{jt},\cV_{jt})$. Note that even if one uses randomized values, we will show that the limiting distributions of our proposed statistics are parameter free and independent of the random vectors $\cV_1,\ldots,\cV_n$. We will also examine the effect of randomization on the finite sample behavior of the proposed tests.

Following \cite{Duchesne/Ghoudi/Remillard:2012}, to test independence between the generalized errors of time series, we will study
the empirical processes based on lagged pseudo-observations.
To deal with more than two time series,
\cite{Duchesne/Ghoudi/Remillard:2012} introduced multivariate lags  $\bfl  = (\ell_1,\ldots,\ell_d) \in \mathbb{Z}^d$ as the
$d$-vector of time lags.
For example, in the classical case $d = 2$, the
vector $\bfl = (0,\ell)$ allows us to consider the usual $l$-dependence between the two
random variables $\ve_{1,t}$ and $\ve_{2,t+\ell}$.
For any $\bfl\in\dZ^d$, we are interested in the limiting distribution of the lagged empirical processes $\dK_{n,\bfl}=\sn(K_{n,\bfl}-\Pi\circ \bF)$, where the empirical joint distribution function of the lagged pseudo-observations  $K_{n,\bfl}$ is defined by
\begin{equation*}
\label{edfH}
K_{n,\bfl}(\by) = n^{-1} \sum_{t=1}^n \I\{\be_{n,t+\bfl} \le \by\}  = n^{-1} \sum_{t=1}^n \prod_{j=1}^d \I \left\{e_{n,j,t+\ell_j} \le y_j \right\}, \; \by \in \dR^d,
\end{equation*}
where the generalized errors $\bve_t$ and the pseudo-observations $\be_{n,t}$  are defined in a circular way, i.e., $\bve_{t+n}=\bve_{t} $ and $\be_{n,t+n}=\be_{n,t} $ for all $t\in\dZ$. Further set $\dF_{n,j} = \sn(F_{n,j}-F_j)$, with $\disp F_{n,j}(y_j) = \frac{1}{n}\sum_{t=1}^n \I\{e_{n,j,t}\le y_j\}$.
To express the asymptotic limit of empirical processes, we also need to define lagged versions of the empirical process of generalized errors, viz.
$\disp
\alpha_{n,\bfl}(\by) =n^{-\frac{1}{2}}\sum_{t=1}^n \left\{
\prod_{j=1}^d \I\{\ve_{j,t+\ell_j}\le y_j\} - \prod_{j=1}^d F_j(y_j)
\right\}, \quad \by \in \dR^d$,
and
$\disp
\beta_{n,\bfl}(\bu) =n^{-\frac{1}{2}}\sum_{t=1}^n \left[
\prod_{j=1}^d \I\{U_{tj}\le u_j\} - \prod_{j=1}^d u_j
\right]$, $\bu \in [0,1]^d$,
where $\bU_t=\bF(\bve_t)$. As a result,
$\alpha_{n,\bfl}=\beta_{n,\bfl}\circ\bF$, for any $\bfl\in\dZ^d$.
Note that circular definition of the errors and of the pseudo-observations makes the formulation of the above processes quite simple and does not alter their asymptotic behaviors since it only involves  a finite number of terms.
Next, to state the first result, one needs additional assumptions listed next. Note that all convergence results hold on the Skorohod spaces $\cD(\bT)$ of c\`ad\l\`ag functions on $\bT$.

For $j\in\setd$, define
$d_{n,jt} = e_{n,jt}-\ve_{jt} -\nabla_{\btheta} \cH_{\btheta_0,jt}(X_{jt},V_{jt})^\top \bTheta_n /\sqrt{n}$, $t\in \setn$, and set
$\bgamma_{jt}(u_j)= E\left\{\nabla_{\btheta} \cH_{\btheta_0,jt}(X_{jt},\cV_{jt})|  U_{jt}=u_j\right\}$. More precisely,
\begin{equation}\label{eq:gamma}
\bgamma_{jt}(u_j)= \left\{\begin{array}{cc} \dot G_{jt}\circ G_{jt}^{-1}(u_j), & \text{ if } \Delta G_{jt}\left\{ G_{jt}^{-1}(u_j)\right\}=0;\\
\dot G_{jt}\left\{ G_{jt}^{-1}(u_j)-\right\} + \Delta \dot  G_{jt}\left\{ G_{jt}^{-1}(u_j)\right\}\left[\frac{u_j-G_{jt}\left\{ G_{jt}^{-1}(u_j)-\right\}}{\Delta G_{jt}\left\{ G_{jt}^{-1}(u_j)\right\}}\right], &\text{ if } \Delta G_{jt}\left\{ G_{jt}^{-1}(u_j)\right\}>0.
\end{array}\right.
\end{equation}
Recall that $\ve_{jt}\sim F_j$, $j\in\setd$.
Then next technical assumptions are adapted from \cite{Ghoudi/Remillard:2018}, \cite{Nasri/Remillard:2019}, and \cite{Nasri/Remillard/Bahraoui:2022},
\renewcommand\theassumption{(A\arabic{assumption})}
\setcounter{assumption}{-1}
\begin{assumption}\label{A0}
There exists $\btheta_0$ so that for any $j\in\setd$ and for any $t>0$, the conditional distribution of $X_{jt}$ given $\cF_{t-1}$ is $G_{\btheta_0,jt}$. Furthermore, under the null hypothesis,
$(\beta_n, \bTheta_{n}) \rightsquigarrow
(\beta, \bTheta)$ in $\cD([0,1]^d)\times \real^p$.
\end{assumption}
\begin{assumption}\label{A1}
For any $j\in \setd$, $\cH_{\btheta,j,t}$  is continuously differentiable with respect to $\btheta \in \cO \subset \real^{p}$, with derivative
denoted by $\nabla_{\btheta} \cH_{\btheta,jt}=\dot \cH_{\btheta,jt}$.
\end{assumption}
\begin{assumption}\label{A2}
  For any $j\in \setd$, there exists a sequence of bounded positive terms $r_{n,jt}>0$ so that
$\disp n^{-\frac{1}{2}}\sum_{t=1}^n r_{n,jt}  $ converges to $0$ as $n\to\infty$
and such that the sequence $\disp
\max_{1 \le t \le n} \|d_{n,jt}\|/r_{n,jt}
$ is tight.
\end{assumption}
\begin{assumption}\label{A3}
    For any $j\in \setd$, $\disp \bGamma_{n,j}(u_j) = \frac{1}{n}\sum_{t=1}^n  \bgamma_{jt}(u_j)
\pr \bGamma_j(u_j)$, and $\bGamma_j(0)=\bGamma_j(1)=0$.
\end{assumption}
\begin{assumption}\label{A4}
 For any $j\in \setd$, $\disp \frac{1}{n}\sum_{t=1}^n E
\left(\|\bgamma_{jt}(U_{jt})\|^k\right) $ is
bounded, for $k=1,2$.
\end{assumption}
\begin{assumption}\label{A5}
For any $j\in\setd$,  $\disp \max_{1\le t\le n}\|\bgamma_{jt}(U_{jt})\|/\sn = o_P(1)$.
 \end{assumption}
 \begin{assumption} \label{A6}
  For any $j\in\setd$, the density  $f_j$ is continuous and bounded in the inside of its support and $\disp
\mathcal{L}_{n,j}(\delta)=\frac{1}{n}\sum_{t=1}^n\sup_{|y_j-x_j|<\delta}\|\gamma_{jt}\circ F_j(y_j)-\gamma_{jt}\circ F_j(x_j)\|f_j(x_j)=o_P(1)$, as $\delta\to 0$.
 \end{assumption}
\begin{remark}\label{rm:hyp}
First, under the null hypothesis, the convergence of $\beta_n$ follows readily. Next, note that the joint convergence in \ref{A0} is usually easy to  prove, especially if $\btheta_n$ is a QMLE or a MLE estimator, In fact, if $\disp \bTheta_n = \sni \sum_{t=1}^n \ell_t+o_P(1)$, where $E(\ell_t|\cF_{t-1})=0$, and $\ell_t$ is stationary and ergodic with finite second moment, then \ref{A0} is satisfied. In particular, $\|\bTheta_n \|$ is tight, so for $\delta>0$, $\disp \sup_{n\ge 1}P(\|\bTheta_n\|>M)<\delta$ if $M$ is large enough.
For interesting results on the estimation of $\btheta_n$ and the convergence of $\bTheta_n$, one might consult \cite{Francq/Zakoian:2010}, \cite{Moysiadis/Fokianos:2014}, \cite{Fokianos/Stove/Tjostheim/Doukhan:2020}, and \cite{Debaly/Truquet:2021}. Note that assumptions \ref{A1}--\ref{A6} must be verified for any $j\in\setd$. First, \ref{A1} depends on the smoothness of the conditional cdf so it is easy to verify.
Now, for many dynamic models, there exists a version  $Y_t$ of $X_{jt}$ which is stationary and ergodic distribution, and in fact the difference  $|Y_t-X_{jt}|$ tends to $0$ exponentially fast.
This is known for many interesting models, including AR-GARCH \citep{Francq/Zakoian:2010}, Markovian regime switching models, including Gaussian HMM used in Section \ref{sec:sim}, ARMA processes and AR-Poisson processes. See, e.g., the Appendix \ref{app:ex}
for a proof in the case of the last two models. For several models, one can also show that there exists an auxiliary process $H_t$ for which $(Y_t,H_t)$ is Markov, with stationary and ergodic initial distribution for $(Y_0,H_0)$. This is the case for Markov regime switching models, where $H_t=\tau_t$ is the regime process, assumed to have a stationary and ergodic distribution. In the case of an AR(1)-GARCH process, $H_t = \sigma_t$ is the volatility. Similar result holds for the  ${\rm INGARCH}(p,q)$ models \citep{Ferland/Latour/Oraichi:2006}. For such models,
 $\bgamma_{t}$ is continuous on $[0,1]$, and is stationary and ergodic as well. If in addition, $\disp \|\bgamma\|_\infty = \sup_{u\in (0,1)} \|\bgamma_{t}(u)\|$ has moments of order $2$, then   \ref{A3}, \ref{A4}, \ref{A6} are satisfied. As a by-product of the continuity and stationarity,  \ref{A5} holds, since for any $\delta>0$,
$$
P\left\{\max_{1\le t \le n}\|\bgamma_{jt}(U_{jt})\| >\delta \right\} \le n P\left\{\|\bgamma_{j1}(U_{j1})\| >\delta \right\} \le \frac{1}{\delta^2}\int_0^1 E\left\{\|\bgamma_{j1}(u)\|^2\
I\left\{\|\bgamma_{j1}(u)\| > \delta \sn\right\}\right\}du \to 0,
$$
as $n\to\infty$.
Next, note that $d_{n,jt} = \frac{1}{2n} \bTheta_n^\top  \left\{\ddot G_{\check\btheta_n,j,t}(X_{jt}-) +\cV_{jt} \Delta \ddot G_{\check\btheta_n,j,t}(X_{jt})\right\}\bTheta_n $, where $\left\|\check\btheta_n -\btheta_0\right\|<  n^{-\frac{1}{2}} \|\bTheta_n\|$. Setting $\disp r_{n,jt} = n^{-1}\sup_{y}\sup_{\|\btheta-\btheta_0\|< n^{-\frac{1}{4}}}\left\|\ddot G_{\btheta,j,t}(y)\right\|$,  and assuming that
$ r_{n,jt} $ is stationary with finite moment, it follows from \ref{A0} that $\ref{A2}$ is satisfied.

The previous conditions can hold for general Stochastic Volatility (SV) models, and in particular for AR-GARCH models,  under stationarity conditions. For a SV model,
$\bgamma_t(u) = -f\circ F^{-1}(u) \bgamma_{0t} - F^{-1}(u) f\circ F^{-1}(u) \bgamma_{1t}$, with $\bgamma_{0t} = \frac{\dot \mu_t}{\sigma_t}$ and $\bgamma_{1t} = \frac{\dot \sigma_t}{\sigma_t}$.
Assuming that $f(y)$ and $yf(y)$ are continuous on $[-\infty,\infty]$, one gets the continuity of $\bgamma_t$ as a function of $u$. Next, the stationarity of $\bgamma_t$ follows generally from the stationarity of  $\bgamma_{0t}$ and $\bgamma_{1t}$.  For example, for an AR-GARCH model
with $\mu_t = \mu+\sum_{j=1}^r \phi_j(X_{t-j}-\mu)$ and
$\disp h_t =\sigma_t^2 = \omega+\sum_{j=1}^q \alpha_j h_{t-j}+\sum_{j=1}^p \beta_j (X_{t-j}-\mu_{t-j})^2 $, $\omega>0$, $\alpha_j,\beta_j\ge 0$, and $\disp \sum_{j=1}^p \beta_j +\sum_{j=1}^q \alpha_j< 1$,
we have
$$
\dot \mu_t = \left(1-\sum_{j=1}\phi_j,X_{t-1}-\mu,\ldots, X_{t-r}-\mu,0,\ldots,0\right)^\top,\mbox{ and}
$$
$$\dot h_t = -2\sum_{j=1}^p \beta_j (X_{t-j}-\mu_{t-j})\dot \mu_{t-j} +  \left(0,\ldots,0,1,h_{t-1},\ldots, h_{t-q},(X_{t-1}-\mu_{t-1})^2,\ldots,(X_{t-p}-\mu_{t-p})^2\right)^\top + \sum_{j=1}^q \alpha_j \dot h_{t-j}.
$$
As a result, all previous conditions are satisfied. It is also possible to satisfy the Assumptions about $\bgamma_t$ even if it is not obvious that $\bgamma_t$ is stationary. In fact, if $G_t$ is stationary and ergodic, so is $G_t^{-1}$ and $\dot G_t$.
In fact,
$$
P\left\{G_{t+1}^{-1}(u) \le y_1,\ldots, G_{t+n}^{-1}(u) \le y_n\right\} = P\left\{G_{t+1}(y_1)\ge u,\ldots, G_{t+n}(y_n)\ge u\right\}= P\left\{G_1(y_1)\ge u,\ldots, G_{n}(y_n)\ge u\right\},
$$
proving that $G_t^{-1}$ is a stationary process.
 Furthermore, $G_t^{-1}(u)$ converges to $G_\infty^{-1}(u)$. If $G_t$ has a bounded continuous density $g_t$, then $\dot G_t$ is Lipschitz and it follows
that for some constant $C$,
$$
\frac{1}{n}\sum_{t=1}\left\|\dot G_t\circ G_t^{-1}(u)- \dot G_t\circ G_\infty^{-1}(u)\right\|\le \frac{C}{n}\sum_{t=1}\left\| G_t^{-1}(u)-  G_\infty^{-1}(u)\right\|\to 0,
$$
as $n\to\infty$. This technique works for example for Gaussian Hidden Markov Models; see, e.g., Appendix \ref{app:hmm}.

Stationary ${\rm INGARCH}(p,q)$ models \citep{Ferland/Latour/Oraichi:2006} also satisfy the previous conditions, since the conditional distribution $G_t$ given $\cF_{t-1}$ is Poisson with parameter $\disp \lambda_t = \omega+\sum_{j=1}^q \alpha_j X_{t-j}+\sum_{j=1}^p \beta_j \lambda_{t-j}$, $\omega>0$, $\alpha_j,\beta_j\ge 0$, and $\disp \sum_{j=1}^p \beta_j +\sum_{j=1}^q \alpha_j< 1$. It then follows that
$\dot G_t(x) = -\dot \lambda_t g_t(x)$, with $g_t(x)=e^{-\lambda_t}\frac{\lambda_t^x}{x!}$, $x\in \{0,1,\ldots\}$, and
$\disp
\dot\lambda_t =(1,x_{t-1},\ldots,x_{t-q},\lambda_{t-1},\ldots,\lambda_{t-p})^\top +\sum_{j=1}^p \beta_j \dot\lambda_{t-j}$.
It follows that $\dot G_t(y)$ is stationary and so is $G_t^{-1}(u)$. In fact,
$$
P\left\{G_{t+1}^{-1}(u) \le y_1,\ldots, G_{t+n}^{-1}(u) \le y_n\right\} = P\left\{G_{t+1}(y_1)\ge u,\ldots, G_{t+n}(y_n)\ge u\right\}= P\left\{G_1(y_1)\ge u,\ldots, G_{n}(y_n)\ge u\right\},
$$
since $(\lambda_{t+1},\ldots,\lambda_{t+n})$ has the same law as $(\lambda_{1},\ldots,\lambda_{n})$.
\end{remark}

Suppose that $\{\beta_\bfl:\bfl\in L\}$ are $d$-dimensional
Brownian bridges, i.e., are continuous centered Gaussian processes with
$\disp
\cov (\beta_\bfl(\bu),\beta_\bfl(\bv))=\prod_{j=1}^d\min(u_j,v_j)-\prod_{j=1}^du_jv_j$,
$\bu, \bv\in [0,1]^d$, so
$\beta^{(1)} (u_1) = \beta_\bfl(u_1,1,\ldots, 1)$, $\ldots$,
$\beta^{(d)}(u_d) = \beta_\bfl(1,\ldots, 1,u_d)$
are
one-dimensional Brownian bridges. Set $\alpha_\bfl = \beta_\bfl \circ \bF$.
\begin{theorem}\label{thm:mainK}
Let $L$ be a finite subset of $\dZ^d$. Then, under the null hypothesis of independence and Assumptions \ref{A0}--\ref{A6},
$\left\{(\beta_{n,\bfl},\alpha_{n,\bfl},\dK_{n,\bfl}); \bfl\in L\right\}$ converge jointly in $\cD[0,1]^d \times \cD[-\infty,+\infty]^{2d}$ to $\left\{(\beta_\bfl,\alpha_\bfl,\dK_\bfl); \bfl\in L\right\}$, where
    $\disp
    \dK_\bfl(\by) = \alpha_\bfl(\by)-\sum_{j=1}^d f_j(y_j) \left(\prod_{k\neq j}F_k(y_k)\right)\bTheta^\top \bGamma_j\circ F_j(y_j), \quad \by\in \dR^d$, and $\bGamma_j$ is defined in Assumption \ref{A3}. In particular,
    setting $\disp
\dF_j(y_j)=\beta^{(j)}\circ F_j(y_j) - f_j(y_j) \bTheta^\top \bGamma_j\circ F_j(y_j)$, $y_j\in \dR$, then for any $j\in\setd$, $\dF_{nj}$ converges in $\cD[-\infty,\infty]$ to $\dF_j$.
\end{theorem}
Next, for any $\bfl\in\dZ^d$, set
$\disp
\cK_{n,\bfl} (\by) =  n^{\frac{1}{2}} \left\{K_{n,\bfl}(\by)-  \prod_{j=1}^d F_{nj} (y_j) \right\}$, $\by\in\dR^d$.
These processes, obtained by replacing $F_j$ with $F_{nj}$ in the definition of $\dK_{n,\bfl}$, can be used to test the null hypothesis of independence. The following interesting result shows that the limiting distribution of $\cK_{n,\bfl}$ does not depend on the estimated parameters.
\begin{cor}\label{cor:mainK}
For any finite subset $L$ of $\dZ^d$, and under the same conditions of  Theorem \ref{thm:mainK},
$\left\{\cK_{n,\bfl}; \bfl\in L\right\}$ converge jointly in $\cD[-\infty,+\infty]^{d}$ to $\left\{\cK_\bfl; \bfl\in L\right\}$, with
$\cK_\bfl=\dC_\bfl\circ \bF$, where
\begin{equation}\label{eq:copulaprocess}
    \dC_\bfl(\bu)=\beta_\bfl(\bu)-\sum_{j=1}^d \beta^{(j)} (u_j)\prod_{k \neq j}  u_k,\quad \bu\in [0,1]^d.
\end{equation}
 \end{cor}
The proof of Corollary \ref{cor:mainK} follows directly from Theorem \ref{thm:mainK} and the multinomial formula; see Appendix \ref{app:proofs} for more details.
In fact, as shown in the proof,
$$ \cK_{n,\bfl}(\by) = \dK_{n,\bfl}(y) - \sn\left\{\prod_{j=1}^d F_{n,j}(y_j)-\prod_{j=1}^d F_j(y_j) \right\}   =
  \dK_{n,\bfl}(y) - \sum_{j=1}^d \left\{\prod_{k\neq j} F_k(y_k)\right\}\dF_{n,j}(y_j) + o_P(1),
  $$
so by Theorem \ref{thm:mainK}, $\cK_{n,\bfl}$ converges in law to
$$
\alpha_{\bfl}(\by) - \sum_{j=1}^d \left\{\prod_{k\neq j} F_k(y_k)\right\} \left\{\beta^{(j)}(F(y_j))-\dF_{j}(y_j)\right\}
 -\sum_{j=1}^d \left\{\prod_{k\neq j} F_k(y_k)\right\} \dF_{j}(y_j) = C_{\bfl}\circ\bF(\by),
$$
showing that $\cK_{n,\bfl}$ has  a parameter free limit. Note that the empirical processes defined by \eqref{eq:copulaprocess} are well known in the literature and are the limit of empirical processes of independent observations. Instead of working with $\cK_{n,\bfl}$, whose limit depends on the unknown marginal distribution functions $F_1,\ldots,F_d$, it is more convenient working with the empirical copula processes
 $\disp
\mathbb{C}_{n,\bfl} (\bu) = \sn \left\{C_{n,\bfl}(\bu)- \prod_{j=1}^d  D_n( u_j )  \right\}$, $\bu\in[0,1]^d$,
where
$\disp
C_{n,\bfl}(\bfu) = n^{-1} \sum_{t=1}^n \prod_{j=1}^d \mathbb{I}\left\{  (n+1)^{-1}R_{n,j,t+\ell_j} \le u_j \right\}$, $ \bfu =(u_1,\ldots,u_d)^{\top}\in [0,1]^d$,
with
\begin{equation}\label{eq:ranks}
R_{n,j,t+\ell_j}=n F_{nj}(e_{n,j,t+\ell_j}), \quad \text{representing the rank of } e_{n,j,t+\ell_j} \quad \text{amongst } e_{n,j,1}, \ldots, e_{n,j,n},
\end{equation}
and
where $D_n$ is the distribution function of the uniform variable over
$\{ (n+1)^{-1}, \ldots, n(n+1)^{-1} \}$, i.e., $D_n(s) = n^{-1} \min\{ n,  \lfloor (n+1)s \rfloor \} = C_{n,\bfl}(s,1,\ldots,1)$,
$s \in [0,1]$. In order to have well-defined ranks for all $t$, we also introduce circular ranks, defined as
$R_{n,j,t} = R_{n,j,t+n}$, $t\in\dZ$, $j\in\setd$.
According to \cite{Genest/Remillard:2004}, using $D_n$ instead of its limit $D$ generally reduces the bias. The proof of the following result follows from Theorem \ref{thm:mainK} and Proposition A.1 in \cite{Genest/Ghoudi/Remillard:2007}.
\begin{cor}\label{cor:maincop}
For any finite subset $L$ of $\dZ^d$, and under the same conditions of  Theorem \ref{thm:mainK},
$\left\{\dC_{n,\bfl}; \bfl\in L\right\}$ converge jointly in $\cD[0,1]^{d}$ to $\left\{\dC_\bfl; \bfl\in L\right\}$.
 \end{cor}

As outlined in \cite{Duchesne/Ghoudi/Remillard:2012} and \cite{Genest/Remillard:2004}, most measures of interdependence are based on the empirical distribution functions $K_{n,\bfl}$ or $C_{n,\bfl}$.
Though, according the above results, both $\mathbb{C}_{n,\bfl}$ and $\cK_{n,\bfl}$  could be used to develop test statistics for testing the
null hypothesis of independence, here, we follow the suggestion of \cite{Ghoudi/Kulperger/Remillard:2001}, \cite{Genest/Remillard:2004}, 
\cite{Duchesne/Ghoudi/Remillard:2012}, and \cite{Nasri/Remillard:2024c}, who argued that tests based on M\"obius transformations of these empirical processes tend to have simpler covariance structures and are more powerful in general. These transformed processes are introduced next.

\subsection{M\"{o}bius transforms}\label{ssec:moebius}
As is \cite{Duchesne/Ghoudi/Remillard:2012}, we introduce useful transformations of $\mathbb{K}_{n,\bfl}$ and $\dC_{n,\bfl}$.
For any $A\subset \cS_d = \{ 1,\ldots,d \}$ and any $\bfl\in\dZ^d$, the M\"obius transformations of the processes
are defined  by
\[
\dK_{n,A,\bfl}(\bfy) = n^{-\frac{1}{2}}
\sum_{t=1}^n \prod_{j \in A}
\left[
       \I \left\{ e_{j,t+\ell_j} \le y_j \right\} -F_{nj}(y_j )
\right],
\]
and
\[
\dC_{n,A,\bfl}(\bfu) = n^{-\frac{1}{2}}
 \sum_{t=1}^n \prod_{j\in A}
\left[
       \I \left\{ R_{j, t+l_j } \le (n+1)u_j \right\}- D_n(u_j)
\right],
\]
with the convention $\disp \prod_{k \in \emptyset} = 1$. Further set
$\disp
\beta_{n,A,\bfl}(\bfu) = n^{-\frac{1}{2}}
\sum_{t=1}^n \prod_{j \in A}\left[  \I \left\{ U_{j,t+\ell_j} \le u_j \right\} -u_j \right]$, $\bu\in [0,1]^d$.

\begin{remark} As noted in \cite{Duchesne/Ghoudi/Remillard:2012}, only sets with $|A|>1$ matters. Furthermore,
we say that $\bfl' \equiv_{A}  \bfl$ if
$l_j-l'_j = l_1- l'_1$ for all $j\in A$.
If $\bfl' \equiv_{A} \bfl$,  then both
$\dK_{n,A,\bfl} - \dK_{A,\bfl',n}$  and  $\dC_{n,A,\bfl}- \dC_{A,\bfl',n}$ converge in probability to $0$. For these reasons, from now on, we only consider finite sets $L$ such that $\ell_1=0$ and sets $A\in \cA = \{A\subset \cS_d; |A|>1\}$.
\end{remark}

We can now describe the asymptotic behavior of these processes.

\begin{theorem}\label{thm:cop}
Let $L$ be a finite subset of $\dZ^d$. Under the same conditions of Theorem \ref{thm:mainK},
$\{ \beta_{n,A,\bfl}, \dC_{n,A,\bfl},\cK_{n,A,\bfl}; A \in\cA, \bfl \in L \}$ converge to continuous centered processes
$\{ \beta_{A,\bfl}, \dC_{A,\bfl},\cK_{,A,\bfl}; A \in\cA, \bfl \in L \}$, where $\dC_{A,\bfl}=\beta_{A,\bfl}$, $\cK_{A,\bfl}=\beta_{A,\bfl}\circ \bF$, and the processes $\beta_{A,\bfl}$ are jointly centered Gaussian, with covariance function
${\Sigma}_{A,\bfl,B,\bfl'}(\bu,\bv)$ between $\beta_{A,\bfl}(\bu)$ and $\beta_{B,\bfl'}(\bv)$, $\bu,\bv \in [0,1]^d$, is
\begin{eqnarray*}
{\Sigma}_{A,\bfl,B,\bfl'}(\bfu,\bfv)&=& \cov \left\{\beta_{A,\bfl}(\bfu),\beta_{B,\bfl'}(\bfv)\right\}
=\left\{ \begin{array}{ll}
      \prod_{j \in A} \left\{ u_j \wedge v_j - u_j v_j \right\}, &
       \mbox{if } B=A  \mbox{ and } \bfl' \equiv_{A} \bfl,\\
         0, & \mbox{otherwise.}
             \end{array}
     \right.
\end{eqnarray*}
\end{theorem}

\subsection{Relationship with \cite{Kheifets/Velasco:2017}}
For any $X$ with distribution function $\cH$ and $\cV\sim \unif$ independent of $X$, set
\begin{equation}\label{eq:khf}
    J_\cH(x,u) = E \left[\I\{\cH(X-)+\cV\Delta \cH(X)\le u\}|X=x\right] = \left\{ \begin{array}{cc}
D\left\{\frac{u- \cH(x-)}{ \Delta \cH(x)}\right\}, & \Delta \cH(x)>0;\\
\I\left\{\cH(x)\le u\right\}, & \Delta \cH(x)=0,\end{array}\right.
\end{equation}
where $D$ is the cdf of a random variable uniformly distributed on $(0,1)$, i.e., $D(u) = \min\left\{1,\max(0,u)\right\}$.
Next, set
$\disp
   \tilde\beta_{n,\bfl}(\bu) =
   E\left\{\beta_{n,\bfl}(\bu)|\bX_1,\ldots,\bX_n\right\}= n^{-\frac{1}{2}}\sum_{t=1}^n \left\{ \prod_{j=1}^d J_{G_{\btheta_0,j,t}}(X_{j,t+l_j},u_j) - \prod_{j=1}^d u_j  \right\}$ and
$\disp
   \tilde\beta_{n,A,\bfl}(\bu) = E\left\{\beta_{n,A,\bfl}(\bu)|\bX_1,\ldots,\bX_n\right\}=n^{-\frac{1}{2}}\sum_{t=1}^n \prod_{j\in A}\left\{ J_{G_{\btheta_0,j,t}}(X_{j,t+l_j},u_j) -  u_j  \right\}$.
\cite{Kheifets/Velasco:2017} used this transform to construct tests of goodness-of-fit and test of serial dependence for univariate discrete time series $X_{jt}$. It is easy to check that not only in the discrete case, but also in the general case, one has
$E\left\{J_{G_{\btheta_0,j,t}}(X_{jt},u_j)|\cF_{t-1}\right\}=u_j$, so for fixed $j$ and  $u_j$, $J_{G_{\btheta_0,j,t}}(X_{jt},u_j)-u_j$ are martingale differences.  In \cite{Kheifets/Velasco:2013}, the authors also showed that when generating $M$ independent copies of uniform variates $\cV_{jt}^{(k)}$, $k\in\{1,\ldots,M\}$, $t\in\setn$, then the average over $k$ of the processes
$$
\beta_{n,k}^{(j)}(u_j) = n^{-\frac{1}{2}}\sum_{t=1}^n \left[ \I\{G_{\btheta_0,j,t}(X_{jt}-)+\cV_{jt}^{(k)}\Delta G_{\btheta_0,j,t}(X_{jt})\le u_j\}-u_j\right],
$$
is an approximation of $\disp \beta_n^{(j)}(u_j)=n^{-\frac{1}{2}}\sum_{t=1}^n \{J_{G_{\btheta_0,j,t}}(X_{jt},u_j)-u_j\}$. In fact, they showed numerically that taking $M=25$ was often enough.  The same procedure can also be applied here to the processes $\beta_{n,\bfl}$, $\beta_{n,A,\bfl}$, $\dK_{n,\bfl}$, $\dK_{n,A,\bfl}$, and $\dK_{n}^{(j)}$, when $F_j=D$. Next, define
$\disp
\tilde \dK_{n,\bfl}(\bu)=n^{-\frac{1}{2}}\sum_{t=1}^n \left\{ \prod_{j=1}^d   J_{G_{\btheta_n,j,t}}(X_{j,t+\ell_j},u_j) - \prod_{j=1}^d u_j  \right\}$,
$\disp
\tilde \dK_{n,A,\bfl}(\bu)=n^{-\frac{1}{2}}\sum_{t=1}^n \prod_{j\in A}\left\{ J_{G_{\btheta_n,j,t}}(X_{j,t+l_j},u_j) -  u_j  \right\}
$,
and
$\disp
\tilde \dK_n^{(j)}(u_j)=n^{-\frac{1}{2}}\sum_{t=1}^n \left\{ J_{G_{\btheta_n,j,t}}(X_{jt},u_j) -  u_j  \right\}$.
We now study  the convergence of the process $\tilde\beta_{n,A,\bfl}$ and $\tilde\dK_{n,A,\bfl}$. To this end, set
$h_{jt}(u_j) = \Delta G_{jt}\circ G_{jt}^{-1}(u_j) $, where $G_{jt}=G_{\btheta_0,jt}$. Next, define $\tilde\bgamma_{jt}=\bgamma_{jt}$.
Further set
\begin{multline*}
\chi_{0,jt}(u,v)= \dI\{h_{jt}(u) >0\}\frac{\left[ G_{jt}\left\{G_{jt}^{-1}(u)\right\} -u\vee v\right] \left[u\wedge v-G_{jt}\left\{G_{jt}^{-1}(u)-\right\}\right] }{ h_{jt}(u)
} \times \dI\left\{G_{jt}^{-1}(u)=G_{jt}^{-1}(v),\right\}.
\end{multline*}
The last two expressions appear in \cite{Kheifets/Velasco:2017} in the context of a univariate discrete time series. Next, to study the asymptotic behavior of processes $\tilde\beta_{n,\bfl}$, $\tilde\beta_{n,A,\bfl}$, $\tilde\dK_{n,\bfl}$, and $\tilde\dK_{n,A,\bfl}$, one needs the following assumptions:
\begin{itemize}
\item[(A5')]
For any $j\in\setd$ and for $\delta>0$ small enough,
$\disp
\frac{1}{n}\sum_{t=1}^n E\left\{ \sup_{y} \sup_{\|\btheta-\btheta_0\|<\delta}\|\dot G_{\btheta,jt}(y)\|^2\right\} =O(1)$.
\item[(A7')] Assume that for any $A$ and any $\bfl\in\dZ^d$, the following limits exist:
$$
\lim_{n\to \infty}\frac{1}{n}\sum_{t=1}^n \prod_{j=1}^d\left\{u_j\wedge v_j-\chi_{0,j,t+\ell_j}(u_j,v_j)\right\}-\prod_{j=1}^d u_j v_j=\kappa_\bfl(\bu, \bv),
$$
and
$$
\lim_{n\to \infty}\frac{1}{n}\sum_{t=1}^n \prod_{j\in A}\left\{u_j\wedge v_j-\chi_{0,j,t+\ell_j}(u_j,v_j)- u_j v_j\right\}=\kappa_{A,\bfl}(\bu, \bv).
$$
\end{itemize}
Next, rename all other assumptions (A0'), \ldots, (A6') when applied to $\tilde\bgamma_{jt}$.

\begin{theorem}\label{thm:khf}
  If Assumptions (A0')--(A7') hold, then, under the null hypothesis of independence, for any finite set $L \subset \dZ^d$,
 $(\tilde\beta_{n,\bfl}, \tilde\beta_{n,A,\bfl} ,\tilde\dK_{n,\bfl}, \tilde\dK_{n,A,\bfl})$ converge jointly in $\ell^\infty \left([0,1]^{4d}\right)$ to $(\tilde\beta_{\bfl}, \tilde\beta_{A,\bfl} ,\tilde\dK_{\bfl}, \tilde\dK_{A,\bfl})$, where $\tilde\dK_{A,\bfl}=\tilde\beta_{A,\bfl}$ and
  $\disp
\tilde\dK_{\bfl}(\bu)=\tilde\beta_\bfl(\bu)-\sum_{j=1}^d \left(\prod_{k\neq j}u_k\right)  \bGamma_{j}(u_j)^\top\bTheta$, where $\bGamma_j$ is defined in Assumption \ref{A3},
  $\tilde\beta_{\bfl}$ is a continuous centered Gaussian processes with covariance function $\kappa_\bfl$,
   while
    $\tilde\beta_{A,\bfl}$ are continuous centered Gaussian processes with
\begin{eqnarray}\label{cov1}
\cov \left\{\tilde\beta_{A,\bfl}(\bfu),\tilde\beta_{B,\bfl'}(\bfv)\right\}
& =& \left\{ \begin{array}{ll}
      \kappa_{A,\bfl}(\bfu,\bfv), &
       \mbox{if } B=A  \mbox{ and } \bfl' \equiv_{A} \bfl;\\
         0, & \mbox{otherwise.}
             \end{array}
     \right.
\end{eqnarray}
In particular, for any $j\in\setd$,
$\tilde \dK_n^{(j)}$ converges in $\ell^\infty\left([0,1]\right)$ to $\tilde \dK^{(j)}$,
where
\begin{equation}\label{eq:kht-univ}
    \tilde \dK^{(j)}(u_j)=\tilde \beta^{(j)}(u_j)-\bTheta^\top \bGamma_j(u_j).
\end{equation}
\end{theorem}
\begin{remark}
First, note that the convergence of $\tilde\dK_n^{(j)}$ in \cite{Kheifets/Velasco:2017} is incorrectly stated. In order to get \eqref{eq:kht-univ}, the $+$ sign in their Equation (4) should be $-$. Next, the distribution of $\tilde\dK_\ell$ is not used to testing, but the distribution of $\tilde\beta_{A,\ell} $ are, and their covariances depend on the unknown distributions $G_{\btheta_0,jt}$ through Assumption (A7'). As a result, the limiting distribution of Cram\'er-von Mises type statistics would be an infinite weighted sum of chi-square distributions, and the weights would depend on these unknown  covariances. See, e.g., \cite[Appendix H]{Nasri:2022} for similar results in another context. The covariances could possibly be estimated by using the estimated functions $G_{\btheta_n,jt}$  and $\dot G_{\btheta_n,jt}$.
\end{remark}
The next section uses the convergence results to develop test statistics for the independence hypothesis studied here.

\section{Test statistics}\label{sec:stats}

Theorems ~\ref{thm:mainK} and \ref{thm:cop} extend the results of \cite{Duchesne/Ghoudi/Remillard:2012} to the context of (extended) generalized error models and show that the limiting behaviour of the empirical processes used to develop independence tests are identical to those obtained for stochastic volatility models. Therefore, all test statistics used in \cite{Duchesne/Ghoudi/Remillard:2012} are applicable in our context and have the same limiting behaviour as the ones in \cite{Duchesne/Ghoudi/Remillard:2012}. All these tests are implemented in the CRAN package IndGenErrors \citep{Ghoudi/Nasri/Remillard/Duchesne:2025}. Here, we will just review those used in our simulation analysis.

\subsection{Cram\'{e}r--von Mises test statistics}
For a given  set
$A \in \mathcal{A}$
and multivariate lag index
$\bfl \in \mathbb{Z}^d$, the Cram\'{e}r--von Mises test statistic is defined by
\begin{eqnarray*}
S_{n,A,\bfl} &=&   \int_{[0,1]^d}\left\{ \dC_{n,A,\bfl}(\bfu) \right\}^2 d\bfu   \label{eq:SnA}\\
                      &=&   \frac{1}{n} \sum_{t=1}^n \sum_{s=1}^n \prod_{j\in A}
                           \left\{
                                 \frac{2n+1}{6n} +\frac{R_{j, t +\ell_j} \left\{R_{j, t +\ell_j}-1
                            \right\}}{2n(n+1)} 
                      \frac{R_{j, s +\ell_j} \left\{R_{j, s +\ell_j}-1\right\}}{2n(n+1)}
                          - \frac{\max\left(R_{j, t +\ell_j},R_{j, s +\ell_j}\right)}{n+1} \right\}.
                          \nonumber
\end{eqnarray*}
Their limiting distributions follow directly from Theorem \ref{thm:cop} and are summarized in the next proposition.
\begin{prop}
Under the conditions given in Theorem~\ref{thm:mainK},
and under the null hypothesis of independence, for any finite set $L\in\dZ^d$,
the sequences
$\{ S_{n,A ,\bfl}$, $A \in \mathcal{A}$, $\bfl \in L\}$
converge in distribution to random variables
$\{ S_{A,\bfl}, A \in \mathcal{A}$, $\bfl \in L \}$
which are defined as $\disp
S_{A,\bfl} = \int_{[0,1]^d} \left\{\beta_{A,\bfl}(\bfu)\right\}^2 d\bfu$.
The random variables
$S_{A,\bfl}$
and
$S_{B,\bfl}$
are independent if
$A \neq B$
or if $
A = B$
and
$\bfl' {\not \equiv}_A \bfl$.
Moreover,
$S_{A,\bfl}= S_{A,\bfl'}$
if
$\bfl' \equiv_{A} \bfl$.
\end{prop}

As pointed out in \cite{Duchesne/Ghoudi/Remillard:2012}, for
any $A \in \mathcal{A}$ of cardinality $|A|$, $S_{A,\bfl}$
has the same distribution as the random variable
$\disp 
    \xi_{|A|} = \sum_{(i_1\ge 1,\ldots, i_{|A|}\ge 1) }\frac{1}{ \pi^{2|A|} (i_1 \cdots i_{|A|})^2} \ Z_{i_1, \cdots, i_{|A|}}^2$,
where $Z_{i_1,\ldots, i_{|A|}}$ are independent $N(0,1)$ random
variables.

\subsection{Generalized cross-correlations} As defined in \cite{Duchesne/Ghoudi/Remillard:2012}, the generalized cross-correlation $\hat r_{A,\bfl}$ is defined by
\begin{equation}
\label{eq:genCC}
\hat{r}_{n,A, \bfl} =    \frac{ \hat{\gamma}_{n,A,\bfl}  }{ \prod_{j\in A} s_{nj}  },
\end{equation}
where $\hat{\gamma}_{n,A,\bfl}=n^{-\frac{1}{2}} \sum_{t=1}^n \prod_{j \in A} \left(  e_{n,j,t+\ell_j} - \bar{e}_j \right)$,
$s^2_{nj} = n^{-1} \sum_{t=1}^n (e_{n,jt}-\bar{e}_j)^2$ and $\bar{e}_j=n^{-1} \sum_{t=1}^n e_{n,jt}$.
Using Theorem~\ref{thm:cop} and arguments similar to \cite{Duchesne/Ghoudi/Remillard:2012},  we obtain that under the null hypothesis,  for any finite set $L\in \dZ^d$, the sequences $\left\{ n^{\frac{1}{2}} \hat{r}_{n,A ,\bfl}, A \in \mathcal{A}, \bfl \in L \right\}$ converge in distribution to standard Normal random variables
$\left\{ r_{A,\bfl}, A \in \mathcal{A}, \bfl \in L \right\} $ with
$\disp
\cov \left(r_{A,\bfl},r_{B,\bfl'}\right) =
\left\{ \begin{array}{ll}
1,  & \mbox{if } A=B  \mbox{ and } \bfl' \equiv_{A} \bfl, \\
0,  &  \mbox{otherwise}.\end{array} \right.$

\subsection{Other dependence measures}\label{ssec:newdepmeas}
Recently, \cite{Nasri/Remillard:2024c} studied the asymptotic behavior of copula-based dependence measures for multivariate data with arbitrary distributions, including their M\"obius transforms.
These include, for example, multivariate extensions of Spearman's rho, Savage's coefficient, and van der Waerden's coefficient.
From now on, let $\bfK = (K_1,\ldots, K_d)$ be a vector of margins with mean $\mu_j$ and finite variance $\sigma_j^2$,  $j\in\setd$, and define
 $\disp \cL_{K_j}(u) = \int_0^u K_j^{-1}(v)dv$.
Next, for any cdf $G$, define
$\disp
\cK_{j,G}(x) = \int_0^1 K_j^{-1}\left\{G(x-)+s \Delta_G(x)\right\}ds$, $j\in \setd$.
Then $\cK_{j,G}(x)  = K_j^{-1}\{ G(x)\} $, if $G$ is continuous at $x$,  and
    $\cK_{j,G}(x)  = \dfrac{\cL_{K_j}\{G(x)\} -\cL_{K_j}\{G(x-)\} }{\Delta_G(x)}$,
if $G$ is not continuous at $x$. The copula-based dependence measures are defined in the following way:\\
For any $A\in\cA$, set
\begin{equation}\label{eq:TnAnonserial}
    \gamma_{\bfK,n,A,\bfl}\left(\widehat C_n^\maltese\right)  = n^{-1}\sum_{t=1}^n \prod_{j\in A}\left\{\cK_{j,F_{n,j}}(e_{n,j,t+\ell_j})-\mu_j\right\}.
\end{equation}
For example, for Spearman's rho, $K_j \equiv  D$, so $\cL_j(u) = {u^2}/{2}$; for van der Waerden's coefficient, $K_j \equiv  \Phi$, so $\cL_j = -\phi\circ \Phi^{-1}$, $\mu_j=0$; for Savage's coefficient, we take $K_j(x) \equiv  e^{x}$, $x\le 0$, so $\cL_j(u) =u\log{u}-u$, $\mu_j=-1$, with the convention that $0\ln{0}=0$. The following result is an immediate consequence of Theorem \ref{thm:cop} and Theorem 1 in \cite{Nasri/Remillard:2024c}.
\begin{cor}\label{cor:cordep}
Under the null hypothesis of independence, for any finite set $L\subset \dZ^d$, $\sn r_{\bfK,n,A,\bfl} =  \sn \dfrac{\gamma_{\bfK,n,A,\bfl}\left(\widehat C_n^\maltese\right)}{\prod_{j\in A} s_{K_j,F_{nj}} }$, $A\in \cA$, converge jointly in distribution to standard Normal random variables
$\left\{ r_{\bfK,A,\bfl}, A \in \mathcal{A}, \bfl \in L \right\} $ satisfying
$\disp \cov \left(r_{\bfK,A.\bfl},r_{\bfK,B,\bfl'}\right) =
\left\{ \begin{array}{ll}
1,  & \mbox{if } A=B  \mbox{ and } \bfl' \equiv_{A} \bfl, \\
0,  &  \mbox{otherwise}.\end{array} \right. $, where
$$
s_{K_j,F_{nj}}^2 = n^{-1}\sum_{t=1}^n \left[\dfrac{\cL_j\{F_{n,j}(e_{n,j,t})\} -\cL_j\{F_{n,j}(e_{n,j,t}-)\} }{\Delta_{F_{n,j}}(e_{n,j,t})}-\mu_j\right]^2 \pr \varsigma_{K_j,F_j}^2, \qquad j\in \setd,
$$
and where, for any cdf $G$, $\disp \varsigma_{K_j,G}^2 = \int \left\{ \cK_{j,G}\{G(x)\} -\mu_j\right\}^2 dG(x)$, $j\in \setd$.
\end{cor}
The latter variances  generally depend on the margins when there are discontinuities; see, e.g., Appendix  \ref{app:dep-measures}.

\subsection{Combining test statistics}
To combine test statistics for different lags and different subsets, we also follow the approach proposed
in \cite{Duchesne/Ghoudi/Remillard:2012}. Clearly, tests for a fixed set or fixed lag will detect a particular type of dependence but will not be consistent for all alternatives. Therefore, combining test statistics with different subsets and different lags could improve the power of the test against a wider range of alternatives.
In the context of tests of independence, following \citet{Littell/Folks:1973},
 \cite{Duchesne/Ghoudi/Remillard:2012}  showed that combing $P$-values produced  powerful test procedures.
To this end, for each subset $A \in \mathcal{A}$, a finite set of multivariate lags $D_A$
can be created so that for any $\bfl,\bfl' \in D_A$ one has $\bfl\,{\equiv}_A\, \bfl'$ if and only if $\bfl=\bfl'$.
The proposed statistics are
\begin{eqnarray}
\label{eq:W}
\cW_n  &=& \sum_{A\in \mathcal{A}} \;\pi^{2(|A|-2)} \sum_{\bfl \in D_A}  \{{\rm S}_{n,A,\bfl}-{\rm B}(n,|A|)\}, \qquad
\cF_n  = -2\sum_{A\in \mathcal{A}} \; \sum_{\bfl \in D_A}\log\left(p_{n,A,\bfl}\right),\\
\cH_n  &=&  n\sum_{A\in \mathcal{A}} \; \sum_{\bfl \in D_A}{\hat r}_{n,A,\bfl}^2 ,\qquad
\cH_{\bfK,n}  = n\sum_{A\in \mathcal{A}} \; \sum_{\bfl \in D_A}{\hat r}_{\bfK,n,A,\bfl}^2 , \label{eq:HK}
\end{eqnarray}
where ${\rm B}(n,|A|)$ is a bias correction term given by
${\rm B}(n,d) = \left(\frac{n-1}{6n}\right)^d - \frac{1}{6^d } + (n-1)\left(\frac{-1}{6n}\right)^d$,
and $p_{n,A,\bfl}=P(\xi_{|A|}>{\rm S}_{n,A,\bfl})$ is the estimated $P$-value of the statistic ${\rm S}_{n,A,\bfl}$.
Under the null hypothesis of independence, the test statistic $\cF_n$ converges in law to a chi-square distribution with
$\disp 2\sum_{A\in \mathcal{A}} |D_A|$ degrees of freedom, while the Wald's statistics $\cH_n$ and $\cH_{\bfK,n}$
converge in law to a chi-square distribution with $\disp \sum_{A\in \mathcal{A}} |D_A|$ degrees of freedom. Finally,
the test statistic $\cW_n$ converges in law to $\disp \sum_{A\in \mathcal{A}} \pi^{2(|A|-2)}\sum_{\bfl \in D_A}{\rm S}_{A,\bfl}$,
under the null hypothesis of independence. \cite{Duchesne/Ghoudi/Remillard:2012} argued that the Edgeworth's expansion, using the first six cumulants, leads to a satisfactory approximation of the $P$-value of $\cW_n$.

Finally, one could define the versions of the previous statistics when limited to pairs. More precisely, set
\begin{eqnarray}
\label{eq:W2}
\cW_{n,2}  &=& \sum_{A\in \mathcal{A}, |A|=2} \;\pi^{2(|A|-2)} \sum_{\bfl \in D_A}  \{{\rm S}_{n,A,\bfl}-{\rm B}(n,|A|)\}, \qquad
\cF_{n,2}  =  -2\sum_{A\in \mathcal{A}, |A|=2} \; \sum_{\bfl \in D_A}\log\left(p_{n,A,\bfl}\right),\\
\cH_{n,2}  &=&  n\sum_{A\in \mathcal{A}, |A|=2} \; \sum_{\bfl \in D_A}{\hat r}_{n,A,\bfl}^2 ,\qquad
\label{eq:HK2}
\cH_{\bfK,n,2} = n\sum_{A\in \mathcal{A}, |A|=2} \; \sum_{\bfl \in D_A}{\hat r}_{\bfK,n,A,\bfl}^2,
\end{eqnarray}
Under the null hypothesis of independence, the test statistic $\cF_{n,2}$ converges in law to a chi-square distribution with
$\disp 2\sum_{A\in \mathcal{A}, |A|=2} |D_A|$ degrees of freedom, while the Wald's statistics $\cH_{n,2}$ and $\cH_{\bfK,n,2}$
converge in law to a chi-square distribution with $\disp \sum_{A\in \mathcal{A}, |A|=2} |D_A|$ degrees of freedom. Finally,
the test statistic $\cW_{n,2}$ converges in law to $\disp \sum_{A\in \mathcal{A}, |A|=2} \pi^{2(|A|-2)}\sum_{\bfl \in D_A}{\rm S}_{A,\bfl}$.

\subsection{Implementation steps}\label{ssec:steps}
Assume we observe a $d$-dimensional times series $\bX_1,\ldots,\bX_n$. Then, use the following steps:
\begin{enumerate}
\item Choose parametric models $G_{\btheta,jt}$, $j\in \setd$, $t\in \setn$. The adequacy of these models can be tested using the tools cited in Remark \ref{rem:gof}, namely \cite{Bai:2003,Kheifets:2015} for continuous cases, and \cite{Kheifets/Velasco:2013,Kheifets/Velasco:2017} for discrete cases.
\item For each $j\in \setd$, estimate $\btheta$ by $\btheta_n$, using for examples the methodologies in   
\cite{Francq/Zakoian:2010}, \cite{Moysiadis/Fokianos:2014}, \cite{Fokianos/Stove/Tjostheim/Doukhan:2020}, and \cite{Debaly/Truquet:2021}.
\item Generate iid random vectors $\bV_1,\ldots,\bV_n$, where $\bV_t = (\cV_{t1},\ldots,\cV_{td})$ are independent and uniformly distributed, and compute the pseudo-observations $e_{n,jt}= \cH_{\btheta_n,jt}(X_{jt},\cV_{jt})$, where
    $$
    \cH_{\btheta_n,jt}(X_{jt},\cV_{jt})=G_{\btheta_n,jt}(X_{jt}-)+\cV_{jt} \left\{G_{\btheta_n,jt}(X_{jt})-G_{\btheta_n,jt}(X_{jt}-)\right\}.
    $$
Furthermore, extend the pseudo-observations $e_{n,jt}$ in a circular way, i.e., $e_{n,j,t+n}=e_{n,jt}$, $t\in\setn$.
\item Compute the statistics $S_{n,A,\bfl}$ (according to Formula \eqref{eq:SnA}, where the ranks are defined in \eqref{eq:ranks}),
$\hat{r}_{n,A, \bfl}$, defined by \eqref{eq:genCC}, and $r_{\bfK,n,A,\bfl}$, defined in Corollary \ref{cor:cordep}.
\item Compute the test statistics $\cW_n,\cF_n,\cH_n,\cH_{\bfK,n}$ according to formulas \eqref{eq:W}--\eqref{eq:HK}. One can also compute
these statistics restricted to pairs, defined respectively by  \eqref{eq:W2}--\eqref{eq:HK2}.
\item The $P$-values of $\cW_n$ can be computed using the Edgeworth's expansion of the first 6 cumulants, while the $P$-values of $S_{n,A,\bfl}$ can be tabulated according to $n$. In addition, the limiting distributions of $\cF_n$, $\cH_n$ and $\cH_{\bfK,n}$ are chi-square distributions with $\disp 2\sum_{A\in \mathcal{A}} |D_A|$ degrees of freedom for $\cF_n$ and $\disp \sum_{A\in \mathcal{A}} |D_A|$ degrees of freedom for $\cH_n$ and $\cH_{\bfK,n}$.
\end{enumerate}
\begin{remark}
Note that Steps 4--6 are implemented in the CRAN package \emph{IndGenErrors} \citep{Ghoudi/Nasri/Remillard/Duchesne:2025}, so one only needs to compute the pseudo-observations $e_{n,jt}$ using Steps 1--3.
\end{remark}

\section{Numerical experiments }\label{sec:sim}
This section presents two sets of numerical experiments. The first set of experiments deals with testing independence between generalized innovations of two time series while in the second set,
we consider the context of three time series.

\subsection{Test of independence between the generalized innovations of two series}\label{sim2d}
For testing for independence between the generalized errors of two time series
$X_{1,t}$ and $X_{2,t}$, $t=1,\ldots,n$, two data generating processes (DGP$_1$ and DGP$_2$) were simulated.
The processes
$\{ X_{1,t} \}$
and
$\{ X_{2,t} \}$
have been chosen so that the degrees of dependence vary between the lagged generalized errors and that the univariate processes are generalized error models.
The two DGPs are described next, after the introduction of some definitions.

\begin{definition}\label{ARX}
A process $\{X_t\}$ is said to follow an AR(p)Z-Gaussian HMM with $J$ regimes and covariates $\bZ_t$ (with first column $\equiv 1$), if the regime at time $t$, $\tau_t\in\{1,\ldots,J\}$, is a finite Markov chain with transition probability matrix  $Q$, such that
$ \disp Q_{j,k} = P(\tau_t=k|\cF_{t-1},\tau_{t-1}=j)$, $j,k\in \{1,\ldots,J\}$, and if the conditional distribution of $X_t$ given $\cF_{t-1}$ and $\tau_t=j$ is Gaussian with mean
$\disp \mu_{t,j}=\sum_{i=1}^p\phi_{i,j}X_{t-i}+\btheta_j^\top \bZ_t$ and  variance $\sigma_j^2$, or equivalently, the distribution of $(X_t-\mu_{t,j})/\sigma_j$ given  $\cF_{t-1}$ and $\tau_t=j$ is  normal with mean $0$ and variance $1$. Here $\cF_{t-1}$ is the information (sigma-algebra) generated by $\{\tau_1,\ldots,\tau_{t-1}, X_1,\ldots,X_{t-1},\bZ_{1}, \ldots, \bZ_t\}$.
Note that, more generally, an AR(p)Z-G HMM is obtained if the distribution of $(X_t-\mu_{t,j})/\sigma_j$ given $\cF_{t-1}$ and $\tau_t=j$ is $G$.

Next, a process $\{X_t\}$ is said to follow an AR(p)Z-Poisson HMM with $J$ regimes and covariate $\bZ_t$ (with first column $\equiv 1$), if the conditional distribution of $X_t$ given $\cF_{t-1}$ and $\tau_t=j$ is a Poisson distribution with mean $\disp \mu_j=\sum_{i=1}^p\phi_{i,j}X_{t-i}+\btheta_j^\top \bZ_t$, with the restriction that all parameters and covariates are non negative.
\end{definition}
The two data-generating processes we used are:\\
\noindent
\textbf{DGP$_1$}: $(u_t,v_{t+\bfl}) \sim C$, where $C$ is a copula in the following copula families: Frank, Gaussian, Clayton (all three with Kendall's tau $\in \{0.1282,1/3\}$), Independence,  and Tent map, with
$X_{1,t} = F_{1t}^{-1}(u_t)$, $X_{2,t} = F_{2t}^{-1}(v_t)$, where $F_{1t}$ and $F_{2t}$ are respectively the conditional distribution of a Gaussian HMM with the parameters given in Table \ref{param1}.
Note that these parameters were chosen to reflect behavior consistent with some stock returns. The values of the Kendall's tau of $\tau=0.1282$ and $1/3$ were chosen, as in \cite{Duchesne/Ghoudi/Remillard:2012}, in order to obtain a correlation coefficient $\rho$ of $0.2$ and $0.5$ respectively, when working with the Gaussian copula.

\begin{table}
\caption{Parameters of the Gaussian HMM used in generating $\{X_{1,t}\}$ and $\{X_{2,t}\}$.}   \label{param1}
\centering
\begin{tabular}{l|c|r|r|rrr}
Series& Regime j& $\theta_{1,j}$&  $\sigma_j$ & $Q_{1,j}$ & $Q_{2,j}$& $Q_{3,j}$ \\ \hline
     &1&  0.002158 & 0.026689 &   0.969080  & 0.030912 & 0.000008\\
$X_{1,t}$&2&  0.004192 & 0.016850 &   0.000233  & 0.169373 & 0.830394 \\
     &3&  0.001306 & 0.008872 &   0.025170  & 0.859641 & 0.115189   \\ \hline
$X_{2,t}$ &1 &  0.000759 & 0.029993 & 0.974388 & 0.025612 \\
     & 2 & 0.000908 & 0.014038 & 0.006381 & 0.993619\\ \hline
\end{tabular}
\end{table}
\noindent
\textbf{DGP$_2$}: $(u_t,v_t)$ are as in DGP$_1$ while
$X_{1,t} = G_t^{-1}(u_t)$ where $G_t$ is a Poisson cdf with mean $\lambda_t=1+0.1 X_{1,t-1}$, and
$X_{2,t} = 0.5 X_{2,t-1}+\eta_t$, with $\eta_t=\Phi^{-1}(v_t)$.

Thus, the process generated under DGP$_1$ corresponds to a continuous generalized errors situation, while
DGP$_2$ is a case where one component is discrete. The null hypothesis of independence is obtained in both cases when the copula $C$ is the independence copula. In our simulation study, for comparison purposes, we considered  $\ell\in\{0,2\}$ and $n\in\{100, 300\}$.
Throughout the simulation and application sections, the combined test statistics $\cW_n$, $\cF_n$, $\cH_n$ and $\cH_{K,n}$ are computed with a maximum lag $M_2=5$ for sets involving two series and a maximum lag $M_3=2$ for sets involving three series.
The statistics $\cH_{K,n}$ are computed for three choices of $K$'s. More precisely, in what follows $\cH_{K,n}$ is denoted $\cH_{S,n}$ when $K$ leads to Spearman's rho, $\cH_{G,n}$  when $K$ leads to van der Waerden's coefficient, and $\cH_{E,n}$ when $K$ defines Savage's coefficient. All simulations use $1000$ Monte-Carlo replicates unless specified differently.
\begin{remark}\label{rem:tent-dep}
    Note that the theoretical values of Kendall's tau, Spearman's rho and van der Waerden coefficient are all $0$  for the independence copula and the Tent Map copula \citep{Remillard:2013}. However, the theoretical Savage's coefficient, defined by  $\rho_E = -1+ E\left\{\log(U)\log(V)\right\}$, $(U,V)\sim C$, is $0$ for the Independence copula, but its value is
$1-\frac{\pi^2}{12}+ \frac{1}{2}\{\log{2}\}^2 \approx 0.41776$  for the Tent Map copula. If, instead, one takes the classical Savage's coefficient, then
$\rho_E = -1+ E\left\{\log(1-U)\log(1-V)\right\} = 1-\frac{\pi^2}{8}\approx -0.2337$ for the Tent Map copula.
\end{remark}

The results for DGP$_1$ are displayed in Table \ref{tab:hmm1} for $\ell=0$. The results for $\ell=2$ are similar so they are not displayed. These results show, in particular, that the empirical levels of the test statistics are quite close to the nominal levels, even for a sample size as low as $n=100$. For all dependence types, the test statistic $\cW_n$ has the best or  close to the best empirical powers, with $\cW_n$ showing slightly better power than $\cF_n$ in all our simulations. Correlation type statistics $\cH_n$, $\cH_{S,n}$, $\cH_{G,n}$ and $\cH_{E,n}$ tend to have a good  power in detecting correlated alternatives, but fail to detect alternatives for which their theoretical value is zero. In particular, among these correlation statistics only $\cH_{E,n}$ has some power in detecting the Tent map alternative. Note it is the only one with non-zero theoretical value under this alternative. As pointed in \cite{Duchesne/Ghoudi/Remillard:2012}, when the dependence structure of the innovations exhibits correlation, the test statistic $\cH_n$ tends to be more powerful than $\cW_n$ and $\cF_n$, confirming the importance of cross-correlations in empirical work. The statistics $\cH_{S,n}$ and $\cH_{G,n}$  have powers quite similar to those of $\cH_n$, while $\cH_{E,n}$ is extremely powerful in detecting Clayton type dependence and much less powerful in detecting Frank type dependence. One also see that $\cH_{G,n}$ is slightly more powerful for detecting Gaussian type dependence. This confirms the finding of \cite{Nasri/Remillard:2024c}, who showed that  among $\cH_{n}$, $\cH_{S,n}$, $\cH_{G,n}$ and $\cH_{E,n}$, the locally most powerful statistics is $\cH_{S,n}$ for Frank's copula,  $\cH_{G,n}$ for the Gaussian copula and $\cH_{E,n}$ for the Clayton copula.

We also note that the power levels for the test statistics $\cW_n$, $\cF_n$ and $\cH_n$ are quite similar to those reported by \cite{Duchesne/Ghoudi/Remillard:2012} for residuals of stochastic volatility models, i.e., the finite sample behavior of our test procedures are quite similar for both continuous errors or generalized error models. Note that, in general, the test based on $\cW_n$ is slightly better than the one based on $\cF_n$, which might be explained by a better approximation of the finite sample distribution. Moreover, the computation of the statistics $\cF_n$ is time consuming and requires many numerical integrations  to evaluate the$P$-values appearing in its definition. Though, theoretically,
 $\cF_n$ represents the optimal way to combine our tests statistics, in practice $\cW_n$ is to be recommended for the simplicity of its computation and for superiority of its performance. The results also confirm that the power  increases within a given copula family as the dependence parameter $\tau$ increases.

\begin{table}[h!]
\caption{
Percentage of rejection of the null hypothesis of independence for DGP$_1$ with $\ell=0$ and test statistics
$\cW_n$, $\cF_n$, $\cH_n$, $\cH_{S,n}$, $\cH_{G,n}$, and $\cH_{E,n}$.
}
\label{tab:hmm1}
\begin{tabular*}{1.0\textwidth}{@{\extracolsep{\fill}}lccccccccc}
\hline
&  & \multicolumn{8}{c}{Copula of $(u_t,v_t)$} \\ \cline{3-10}
$n$ & Test statistics  &Independence &\multicolumn{2}{c}{Frank}&\multicolumn{2}{c}{Normal}&\multicolumn{2}{c}{Clayton}&Tent map\\
$\tau$ & &  & 0.1282& 1/3& 0.1282& 1/3 & 0.1282& 1/3& \\
\hline
100& $\cW_n$  & 4.5 & 16.4 & 87.6 & 12.2 & 85.6  & 13.9 & 88.3 & 91.5         \\
&  $\cF_n$    & 4.4 & 14.0 & 81.9 & 10.2 & 80.1  & 12.2 & 83.2 & 83.5      \\
& $\cH_n$     & 5.1 & 17.1 & 90.1 & 14.2 & 91.3  & 16.5 & 91.2 & 5.5       \\
& $\cH_{S,n}$     & 4.9 & 17.5 & 89.8 & 14.0 & 90.8  & 16.8 & 91.0 & 4.9          \\
& $\cH_{G,n}$     & 4.8 & 16.3 & 88.6 & 14.9 & 92.8  & 18.1 & 93.1 & 4.3      \\
& $\cH_{E,n}$     & 5.4 & 12.5 & 68.2 & 14.0 & 80.6  & 32.5 & 97.8 & 47.9      \\
\hline
300& $\cW_n$  & 4.2 &  50.9 &  100 & 47.1 & 100 & 47.0 & 100  & 100      \\
&  $\cF_n$    & 4.1 &  45.7 &   99.8 & 43.1 & 99.9  & 44.1 & 100   & 100    \\
& $\cH_n$     & 4.4 &  54.2 &  100 & 52.4 & 100 & 51.1 & 100  & 7.6     \\
& $\cH_{S,n}$     & 4.7 &  53.8 &  100 & 52.7 & 100 & 50.2 & 100  & 7.7        \\
& $\cH_{G,n}$     & 5.2 &  49.2 &  100 & 56.0 & 100 & 58.6 & 100  & 5.5     \\
& $\cH_{E,n}$     & 5.2 &  32.2&   100 & 40.0 & 99.9 & 82.2 & 100  & 96.1    \\
\hline
\end{tabular*}
\end{table}


By definition, tests based on $\cW_n$, $\cF_n$, $\cH_{S,n}$, $\cH_{G,n}$ and $\cH_{E,n}$ should not depend on the marginal distribution of the errors/generalized errors. However, tests based on $\cH_n$ should inherit cross-correlations properties and are expected to be more powerful for Gaussian innovations and for linear dependence.
To explore dependence on the marginal distributions, we selected DGP$_1$,   but instead of AR(p)Z-Gaussian HMM, we used AR(p)Z-G HMM, with $G$ being a centered standard exponential distribution or a centered Pareto distribution with tail parameter equal to six, that is, the distribution is given by $G(x)=1-(x+6/5)^{-6}$, for $x \ge -1/5$. More precisely, for $t=1,\ldots, n$ the couple $(u_t,v_{t})$ are generated from a Gaussian copula with Kendall's tau $\tau \in \{0,0.1283, 1/3\}$ and $X_{1,t}$ and $X_{2,t}$ are generated according to an AR(p)Z-centered Exponential or centered Pareto HMM with parameters given in Table \ref{param1}.
The results displayed in Table \ref{tab:hmm3}, for the sample sizes $n \in\{100, 300\}$, show  that  when comparing Gaussian marginals to Exponential marginals, note that for $n=300$, $\cW_n$ and $\cF_n$ have similar powers in both cases, while a significant reduction of power is observed for Pareto's margins. When $n=100$, and specially for weak dependence, the power of all statistics is slightly better under Exponential margins, for all but the statistic $\cH_{E,n}$. On the other-hand, the power for Pareto marginals is clearly smaller than that for the other margins, regardless of the statistics, the sample size, or the dependence strength.
 One also notes that the drop in power of $\cH_n$ is surprisingly lower than the one observed in \cite{Duchesne/Ghoudi/Remillard:2012}, while the drop in  power of $\cW_n$ and $\cF_n$, for Pareto marginals, is clearly higher in our case compared to \cite{Duchesne/Ghoudi/Remillard:2012}.  The difference in behavior is due to the fact that test statistics are applied to the residuals in \cite{Duchesne/Ghoudi/Remillard:2012}, while here, the test statistics are applied to generalized errors, which are already transformed to have uniform distribution for all choices of marginal distributions.

 \begin{table}[h!]
 \caption{Rejection levels of the test statistics
 $\cW_n$, $\cF_n$ and $\cH_n$ under DGP$_1$ (HMM-3, HMM-2) with different marginal distributions and with $\ell=0$. }\label{tab:hmm3}
 \begin{tabular*}{1.0\textwidth}{@{\extracolsep{\fill}}ccccccccccc}\hline
  &             &\multicolumn{9}{c}{Marginal distributions} \\ \cline{3-11}
  $n$ & Test statistics    &\multicolumn{3}{c}{Gaussian}   &\multicolumn{3}{c}{Exponential}&\multicolumn{3}{c}{Pareto}\\ \cline{3-5}\cline{6-8} \cline{9-11}
 $\tau$ & &Indep.& 0.1282& \multicolumn{1}{c}{1/3 }& Indep.& 0.1282& \multicolumn{1}{c}{1/3 }& Indep.& 0.1282&\multicolumn{1}{c}{1/3 }\\  \hline
 100 &$\cW_n$& 4.5& 12.2 & 85.6& 5.0& 14.2   &   86.4   &5.0&   9.2   &   58.5 \\
     &$\cF_n$& 4.4& 10.2& 80.1& 4.8& 13.1   &   80.3   &  4.8&  7.7   &   52.0 \\
     &$\cH_n$& 5.2& 14.2& 91.3&4.7& 17.8   &   91.2   & 5.4&   10.2   &   63.9 \\
     &$\cH_{S,n}$ & 4.9& 14.0 & 90.8&4.6& 17.6 &   91.5 & 5.1&  10.2   &   64.6 \\
     &$\cH_{G,n}$& 4.8& 14.9 & 92.8& 4.5&18.7 &   92.8 &  5.2&  10.2   &   66.1 \\
     &$\cH_{E,n}$& 5.4&  14.0 & 80.6&5.8& 14.0 &   73.7 &  5.7&  8.9   &   38.1 \\
     \hline
 300 &$\cW_n$& 4.2&47.1& 100& 5.0&46.5  &   100    &   6.2& 29.9   &   99.1  \\
     &$\cF_n$& 4.1&43.1& 99.9& 5.1& 43.3   &   99.7    &  5.9&  27.2   &   98.3   \\
     &$\cH_n$& 4.4&52.4& 100& 4.5& 45.3   &   100   &   6.8& 31.0  &   99.2   \\
     &$\cH_{S,n}$&4.7& 52.7& 100& 4.0& 49.4   &   100   & 5.7&   30.8   &  99.5 \\
     &$\cH_{G,n}$& 5.2&56.0& 100& 4.7& 51.3  &   100   &  6.0&  32.9  &   99.6 \\
     &$\cH_{E,n}$& 5.2&40.0& 99.9&7.8& 31.8   &   99.8   & 8.9&  21.0   &   88.1\\
     \hline
 \end{tabular*}
 \end{table}

Next, to assess the effect of discrete components among the time series, we generated the time series under DGP$_2$ with different copulas and dependence parameters and applied our testing procedures. Since the conditional distribution of the first component is discrete, the testing procedure requires randomization through the vector $\cV_t$ given in Equation \eqref{eq:extge}. Table~\ref{tab:dgp2} displays the results for sample sizes $n\in\{100, 300\}$. By comparing with Table~\ref{tab:hmm1} or with Table 1 in \cite{Duchesne/Ghoudi/Remillard:2012}, we notice that the powers of all test statistics are a bit lower than those of the continuous generalized errors cases.
Though the loss in power is not substantial in our context,  \cite{Kheifets/Velasco:2017} noted that such reduction results from the randomization introduced by $\cV_t$. They argued that the problem could be eliminated or reduced by integrating with respect to the distribution of $\cV_{jt}$ or by averaging over a large number of randomizations. Here, we explore the effect of randomization by considering averaging over $M$ randomizations for different values of $M\in \{1, 25,50, 100\}$. For $M=1$, our statistics are distribution-free under the null hypothesis of independence, while for $M >1$, for each randomization $k=1,\ldots, M$,  the test statistics has a distribution free limit, but these limits are not independent for different $k$'s. Therefore,  the averaged statistics are no longer distribution-free, and its limiting distribution depends on the dependence structure between the statistics for different randomizations. To obtain the distribution under the null hypothesis of independence,  we simulated data from DGP$_2$ with the Independence copula, and for each simulation iteration, we considered $100$ random randomizations $\cV_{jt}^{(k)}$ for $k\in \{1,\ldots, 100\}$. For each randomization $k$, we computed the values of the statistics $\cW_n^{(k)}$, $\cF_n^{(k)}$, $\cH_n^{(k)}$, $\cH_{S,n}^{(k)}$, $\cH_{G,n}^{(k)}$ and $\cH_{E,n}^{(k)}$,  and we calculated the average of the first $M$ of these statistics for  $M \in\{25, 50, 100\}$. We performed $10,000$ replications, and we computed the quantiles under the null hypothesis of independence. The results for the $95$th and $99$th quantiles are displayed in Table~\ref{quant}. One can see that the results are quite stable as soon as $M\ge 25$, and there is no huge difference between the quantiles for $M=1$ and $M=25$.

\begin{table}[h!]
\caption{
Percentage of rejection of the null hypothesis of independence for DGP$_2$ with $\ell=0$ and test statistics
$\cW_n$, $\cF_n$, $\cH_n$, $\cH_{S,n}$, $\cH_{G,n}$, and $\cH_{E,n}$.
}
\label{tab:dgp2}
\begin{tabular*}{1.0\textwidth}{@{\extracolsep{\fill}}lccccccccc}
\hline
&  & \multicolumn{8}{c}{Copula of $(u_t,v_t)$} \\ \cline{3-10}
$n$ & Test statistics  &Independence &\multicolumn{2}{c}{Frank}&\multicolumn{2}{c}{Normal}&\multicolumn{2}{c}{Clayton}&Tent map\\
$\tau$ & & 0 & 0.1282& 1/3& 0.1282& 1/3 & 0.1282& 1/3& 0\\
\hline
100& $\cW_n$      & 3.6 & 12.5 & 82.4 & 11.7 & 80.2 & 11.9 & 79.8 & 81.4        \\
&  $\cF_n$        & 3.1 & 10.8 & 77.8 & 10.3 & 73.0 & 10.7 & 72.9 & 69.8      \\
& $\cH_n$         & 4.2 & 14.2 & 84.8 & 13.2 & 88.6 & 17.7 & 83.8 & 7.3       \\
& $\cH_{S,n}$     & 4.0 & 13.8 & 86.8 & 12.8 & 86.3 & 16.0 & 82.1 & 6.4          \\
& $\cH_{G,n}$     & 4.9 & 13.3 & 83.0 & 11.8 & 87.8 & 17.2 & 80.1 & 7.0       \\
& $\cH_{E,n}$     & 5.3 & 9.9 & 57.5 &  10.1 & 59.5 & 26.4 & 76.4 & 8.3       \\
\hline
300& $\cW_n$       & 4.5  & 43.6 & 100    & 41.7 & 100 & 44.3 & 100 & 100         \\
&  $\cF_n$         & 4.6  & 39.5 & 99.8   & 37.6 & 99.8 & 40.0 & 99.9 & 99.6         \\
& $\cH_n$          & 4.4  & 45.2 & 100    & 47.3 & 100 & 55.6 & 100 & 10.2            \\
& $\cH_{S,n}$      & 4.6  & 47.2 & 100    & 45.6 & 100 & 47.8 & 100 & 7.4           \\
& $\cH_{G,n}$      & 4.2  & 41.7 & 100    & 45.5 & 100 & 54.4 & 100 & 14.5       \\
& $\cH_{E,n}$      & 6.2  & 25.1 & 98.6   & 25.5 & 99.3 & 74.2 & 100 & 16.0       \\
\hline
\end{tabular*}
\end{table}

\begin{table}
\caption{Quantiles of the null distribution under DGP$_2$ for different values of $M$, when $n=100$. Results are based on $10,000$ Monte-Carlo iterations.}
\label{quant}
\centering
\begin{tabular}{c|cc|cc|cc|cc}
\hline
$M$&\multicolumn{2}{c|}{1}& \multicolumn{2}{c|}{25} &\multicolumn{2}{c|}{50} & \multicolumn{2}{c}{100}\\ \hline
& $q_{95}$& $q_{99}$& $q_{95}$& $q_{99}$&$q_{95}$& $q_{99}$ &$q_{95}$& $q_{99}$ \\ \hline
$\cW_n^M$ & 0.3943 & 0.4421 & 0.3848  & 0.4300 & 0.3841 & 0.4288 & 0.3839 & 0.4297  \\
$\cF_n^M$ & 33.2000 & 38.9599 & 31.8923 & 37.3231 & 31.8496 & 37.1933 & 31.8636 & 37.2346  \\
$\cH_n^M$ & 19.4536 & 24.3266 & 18.7266 & 23.1122 & 18.7172 & 23.0265 & 18.7439 & 23.0188  \\
$\cH_{S,n}^M$ &19.5433 & 24.2022 & 18.6550 & 22.9820 & 18.6609 & 23.0572 & 18.6919 & 23.0834\\
$\cH_{G,n}^M$ &19.5206 & 24.4913 & 18.4043 & 22.7264 & 18.3214 & 22.9112 & 18.3996 & 22.7888 \\
$\cH_{E,n}^M$ &20.0694 & 24.9714 & 17.0520 & 20.2014 & 17.0177 & 20.3326 & 16.9693 & 20.2008  \\
\hline
\end{tabular}
\end{table}

To study the behavior of the averaged statistics, we generated data according to DGP$_2$ with a Gaussian copula having $\tau=0.1283$ or $1/3$. The results of $1,000$ Monte-Carlo simulations are displayed in Table~\ref{aver1}. Clearly, averaging improved the power of the test statistics.
There was an improvement in the power for $M=1$, which is due to the fact that in this simulation experiment, we used the empirical quantiles of Table \ref{quant}, rather than the asymptotic distributions used in all other experiments. Also note that the power of the averaged tests is similar to that of continuous generalized errors, even when $M=25$.

\begin{table}[h!]
 \caption{Percentage of rejection for different values of $M$ for data generated under DGP$_2$ with Gaussian copula with $\tau\in \{0.1283, 1/3\}$ and $n=100$.}
 \label{aver1}
\begin{center}
\begin{tabular}{c|cccc|cccc}
\hline
& \multicolumn{8}{c}{Gaussian copula}\\
& \multicolumn{4}{c|}{$\tau=0.1283$}  & \multicolumn{4}{c}{$\tau=1/3$} \\ \cline{1-9}
& \multicolumn{4}{c|}{M}  & \multicolumn{4}{c}{M} \\
& 1 & 25 & 50 & 100       &  1 & 25 & 50 & 100 \\ \hline
$\mathcal{W}_n$&15.2 &16.5& 17.0 & 17.0  &    82.8 & 87.4 & 87.3 & 87.4  \\
$\mathcal{F}_n$&14.6 &16.2& 16.3 & 16.5  &    77.6 & 83.1 & 83.2 & 83.5 \\
$\mathcal{H}_n$&17.7 &17.1& 17.6 & 17.6  &    90.3 & 92.6 & 92.7 & 92.7  \\
$\mathcal{H}_{S,n}$&16.4 &18.5 &18.3 &18.2 &    88.3 & 91.3 & 91.6 & 91.2  \\
$\mathcal{H}_{G,n}$&15.6 &17.8 &18.2 &18.1  &    89.3 & 93.1 & 93.2 & 92.9   \\
$\mathcal{H}_{E,n}$&11.3 &14.8 &14.7 &15.3  &    58.5 & 82.0 & 82.5 & 81.9  \\
\hline
\end{tabular}
\end{center}
\end{table}

\subsection{Test of independence between the generalized innovations of three time series}\label{sim3d}

For testing the independence between three time series
$\{X_{1,t}\}$, $\{X_{2,t}\}$ and $\{X_{3,t}\}$, $t\in \setn$, we generated data according to the following  data generating process DGP3.\\
\noindent
\textbf{DGP$_3$}: Generate $(u_t,v_{t},w_t)$ following a copula $C$ and define
$X_{1,t}=G_t^{-1}(u_t)$ where $G_t$ is the cdf of a Poisson random variable with mean $\lambda_t=1+0.1X_{1,t-1}$,
$X_{2,t} = 0.5X_{2,t-1}+\ve_t$, where $\ve_t=\Phi^{-1}(v_t)$, with $\Phi$ the cdf of the standard normal distribution, and
$X_{3,t} = 0.5X_{3,t-1}+\eta_t$, where $\eta_t=\Phi^{-1}(w_t)$.
The copula $C$ is either the Independence copula, a Frank copula with
Kendall's tau $\in \{0.1283,1/3\}$, a Gaussian copula with
Kendall's tau $\in \{0.1283,1/3\}$,  a Clayton copula with Kendall's tau $\in \{0.1283,1/3\}$ or
Romano's copula \citep{Romano/Siegel:1986} associated with $(U,V,W)$,
$\disp W=\left\{\begin{array}{ll} \eta & \mbox{ if } (U-\frac{1}{2})(V-\frac{1}{2})(\eta-\frac{1}{2})\ge 0 \\
1-\eta & \mbox{ if }  (U-\frac{1}{2})(V-\frac{1}{2})(\eta-\frac{1}{2}) < 0
\end{array}\right.$,
where $(U,V,\eta)$ are independent uniform $(0,1)$ random variables. Note that  $(U,V,W)$ are pairwise independent but not jointly independent.

The simulation study was based on $1000$ Monte-Carlo iterations. For all statistics, we considered all sets $A$ with two or three series. For sets with two series, the maximum lag was $M_2=5$, while for sets involving three series, the maximum lag was $M_3=2$.
We also considered the statistics $\cW_{n,2}$, $\cF_{n,2}$, $\cH_{n,2}$, $\cH_{S,n,2}$, $\cH_{G,n,2}$, and $\cH_{E,n,2}$, restricted to sets of cardinality 2, and defined by \eqref{eq:W2}--\eqref{eq:HK2}.
Table~\ref{disc3d1} reports the simulation results. It shows, in particular,  that the empirical levels under the null hypothesis of independence are reasonably close to the nominal level of $5\%$ for all statistics but $\cH_{E,n}$, where the level is around $8\%$ or $9\%$. To investigate this, we generated i.i.d. uniforms random  vectors and we computed the statistic $\cH_{E,n}$. Table \ref{tab:savage} displays the level and the $95$th quantiles for different sample sizes. These results show that this statistic converges quite slowly to its asymptotic distribution when $d=3$, while it converges quite fast for $d=2$. Returning to Table~\ref{disc3d1}, we can see that the empirical power of $\cW_n$ and $\cF_n$ is slightly lower for the Gaussian copula   than for for Clayton or Frank copulas when $n=300$ and $\tau=0.1282$. Also, for $n=300$, $\cH_{S,n}$ is the best for Frank's copula, while $\cH_{G,n}$ is the best for Gaussian copula, and $\cH_{E,n}$ is the best for Clayton's copula.
$\cW_n$ seems to be a bit more powerful for the Romano-Siegel copula when $n=100$, but for $n=300$, the statistics $\cW_n$, $\cF_n$, $\cH_n$, $\cH_{S,n}$ and $\cH_{G,n}$ have similar power, while the power of $\cH_{E,n}$ lags behind. None of the statistics based on sets of cardinality two is able to detect Romano's type dependence, so these statistics are not recommended.   Compared with the results in Table 7 of \cite{Duchesne/Ghoudi/Remillard:2012},  we observe a slight loss of power for the all statistics, similar to the loss of power observed for DGP$_2$. This is attributed, as before, to the randomization effect and could be reduced by applying the averaging technique used in Section \ref{sim2d}.

\begin{table}[h!]
\caption{
Percentage of rejection of the null hypothesis of independence for DGP$_3$ with $\ell=0$ and test statistics
$\cW_n$, $\cF_n$, $\cH_n$, $\cH_{S,n}$, $\cH_{G,n}$, $\cH_{E,n}$, $\cW_{n,2}$, $\cF_{n,2}$, $\cH_{n,2}$, $\cH_{S,n,2}$, $\cH_{G,n,2}$, and $\cH_{E,n,2}$.
}
\label{disc3d1}
\begin{tabular*}{1.0\textwidth}{@{\extracolsep{\fill}}lccccccccc}
\hline
&  & \multicolumn{8}{c}{Copula of $(u_t,v_t,w_t)$} \\ \cline{3-10}
$n$ & Test statistics  &Independence &\multicolumn{2}{c}{Frank}&\multicolumn{2}{c}{Normal}&\multicolumn{2}{c}{Clayton}&Romano-Siegel\\
$\tau$ & & 0 & 0.1282& 1/3& 0.1282& 1/3 & 0.1282& 1/3& 0\\
\hline
100& $\cW_n$       &4.1 & 18.1 & 97.7 & 13.5 & 94.8 & 18.9 & 95.7 & 83.8 \\
    & $\cF_n$      &2.7 & 12.9 & 94.4 & 9.9 & 88.3 & 12.9 & 90.9 & 66.1 \\
 & $\cH_n$         &5.7 & 23.3 & 99.2 & 21 & 99 & 33.9 & 98.1 & 57.3 \\
& $\cH_{S,n}$     &5.3 & 24.9 & 99.1 & 18.6 & 98.1 & 22.5 & 97.6 & 79.3 \\
& $\cH_{G,n}$     &5.2 & 23 & 98.9 & 19.9 & 98.8 & 28.9 & 97.5 & 53.8 \\
& $\cH_{E,n}$     &8.9 & 21 & 85.5 & 22.6 & 88.4 & 57.9 & 99.3 & 26.8 \\
& $\cW_{n,2}$     &4.2 & 25.6 & 99.5 & 22 & 98.7 & 24.5 & 99.2 & 4.4 \\
& $\cF_{n,2}$     &4.1 & 22.5 & 99 & 19.6 & 97.4 & 21.1 & 98 & 4.1 \\
& $\cH_{n,2}$     & 5.6 & 29.5 & 99.6 & 30.3 & 99.9 & 37 & 99.7 & 4.7 \\
& $\cH_{S,n,2}$   &5.1 & 29.7 & 99.6 & 28.3 & 99.8 & 29.5 & 99.4 & 4.2 \\
& $\cH_{G,n,2}$   &5.7 & 27.7 & 99.6 & 29.5 & 99.8 & 34.3 & 99.2 & 4.6 \\
& $\cH_{E,n,2}$   & 5.3 & 18.3 & 94.9 & 19.6 & 95.1 & 50.4 & 99.9 & 4.2 \\
\hline
300& $\cW_n$       &4.8 & 67.7 & 100 & 61.5 & 100 & 66 & 100 & 100 \\
& $\cF_n$         &4.6 & 64.1 & 100 & 57.2 & 100 & 62.5 & 100 & 100 \\
& $\cH_n$         &4.9 & 70 & 100 & 76.2 & 100 & 83.2 & 100 & 100 \\
& $\cH_{S,n}$     &4.6 & 74.2 & 100 & 71.3 & 100 & 72.9 & 100 & 100 \\
& $\cH_{G,n}$     & 5.3 & 68.1 & 100 & 75.4 & 100 & 81.2 & 100 & 100 \\
& $\cH_{E,n}$     &8.4 & 43.4 & 100 & 54.1 & 100 & 96.2 & 100 & 73.6 \\
& $\cW_{n,2}$     &3.8 & 78.3 & 100 & 77.3 & 100 & 76.8 & 100 & 5.0 \\
& $\cF_{n,2}$     &3.7 & 74.4 & 100 & 74.1 & 100 & 74.3 & 100 & 5.1 \\
& $\cH_{n,2}$     &4.5 & 79.7 & 100 & 86.7 & 100 & 89.4 & 100 & 4.8 \\
& $\cH_{S,n,2}$   & 4 & 83.8 & 100 & 84.7 & 100 & 81.6 & 100 & 4.4 \\
& $\cH_{G,n,2}$   &5.3 & 78.1 & 100 & 85.9 & 100 & 87.6 & 100 & 5.4 \\
& $\cH_{E,n,2}$   &5.7 & 53.5 & 100 & 64.8 & 100 & 97.7 & 100 & 5.8 \\
 \hline
 \end{tabular*}
\end{table}
\begin{table}[h!]
 \caption{Percentage of rejection and quantiles for $\cH_{E,n}$ in the d-dimensional case, with $d=2$ or $3$, $M_2=5$ and $M_3=2$, using B=100,000 iterations of i.i.d. uniforms. The empirical level for $n=\infty$ is the target 5\% and the quantile for $n=\infty$ is the 95\% quantile of the limiting Chi Square distribution with $11$ degrees of freedom for $d=2$ and $58$ degrees of freedom for $d=3$.}
 \label{tab:savage}
 \centering
 \begin{tabular}{cccccccc}
 \hline
& & \multicolumn{6}{c}{$n$} \\
 \cline{3-8}
$d$ & &  100 & 300 & 500 & 750 & 1000 & $\infty$ \\
 \hline
2& Empirical level &5.147& 5.255& 4.993& 4.983& 5.167& 5.0 \\
& $95\%$ quantile &19.78523& 19.84092& 19.67329& 19.66369& 19.791& 19.67514\\ \hline
3& Empirical level &9.782 &8.444 &7.686 & 6.961 &6.827 &5.0 \\
& $95\%$ quantile &84.65353 &82.00905 &80.46056 &79.55785 &79.18585 &76.7778 \\
 \hline
  \end{tabular}
 \end{table}

Overall, one sees that when working with continuous generalized errors, the behavior of our tests statistics are almost identical to those reported in \cite{Duchesne/Ghoudi/Remillard:2012} for continuous innovations. When discrete or non-continuous generalized errors are considered, then applying our test procedures with a single randomization has the advantage of producing distribution-free test statistics at the expense of a small loss of power. Averaging over $M$ randomizations improves the power to levels similar to those observed for continuous generalized errors, but this makes the statistics lose their distribution-free property and forces the users to implement a computationally intensive re-sampling technique such as parametric bootstrap. One also notes that the statistics $\cW_n$ is powerful in detecting all types of alternatives, while other statistics might fail in detecting certain types of dependence.

\section{Applications}\label{sec:applications}

We consider two applications. The first one, presented in Section \ref{ssec:nasdaq}, deals with stock returns and demonstrates the use of our testing procedures for zero-inflated generalized errors framework.
The second application, described in Section \ref{ssec:crime}, deals with (integer-valued) Pittsburgh crime time series and
illustrates our methodology for discrete generalized errors.

\subsection{Application for financial data from the NASDAQ  market}\label{ssec:nasdaq}

\begin{figure}[h!]
  \begin{center}
\includegraphics[scale=0.25]{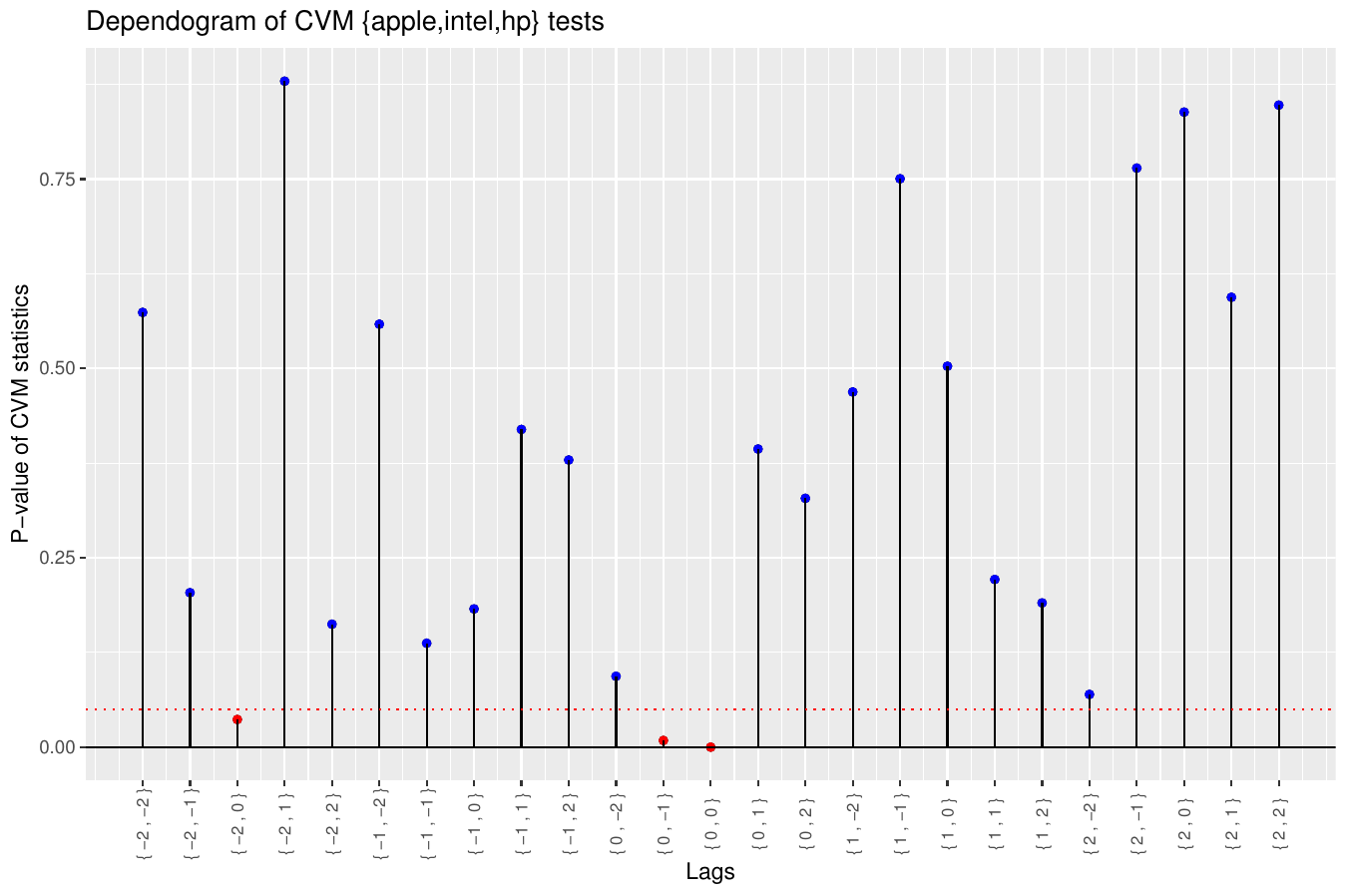}
\includegraphics[scale=0.25]{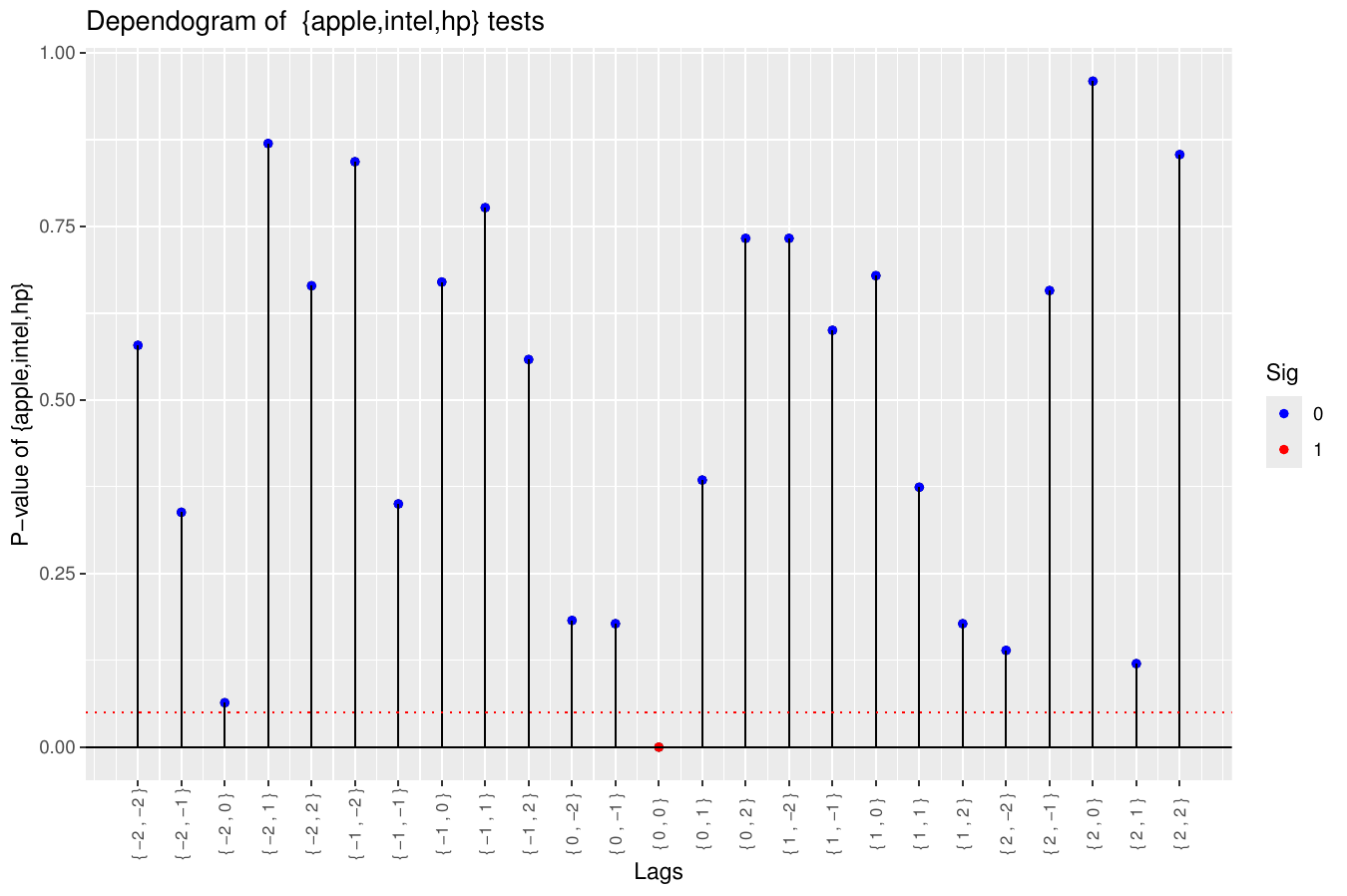}
\end{center}
\caption{Dependogram of Cram\'er-von Mises statistics for the  Apple, Intel and Hewlett-Packard returns (left panel) and their generalized errors with no covariates (right panel).}
\label{dep0}
\end{figure}

\begin{figure}[h!]
  \begin{center}
\includegraphics[scale=0.25]{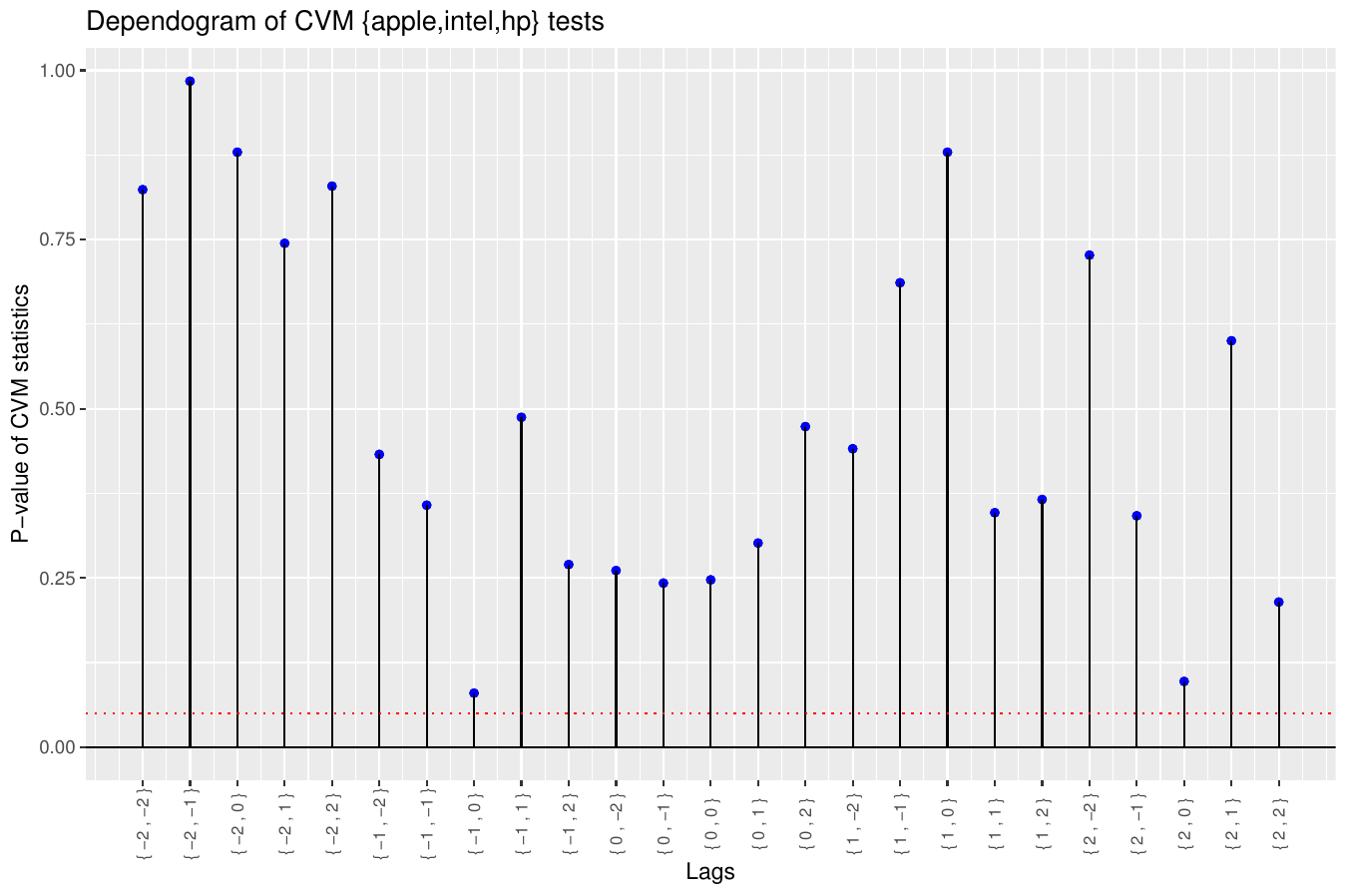}
\end{center}
\caption{Dependogram of Cram\'er-von Mises statistics for the generalized errors of Apple, Intel and Hewlett-Packard AR-Gaussian HMM models, including NASDAQ as a covariate.}
\label{dep123}
\end{figure}

We consider the data set used by \cite{Duchesne/Ghoudi/Remillard:2012}, which consists of log-returns for  Apple, Intel and
Hewlett-Packard,  January 2nd, 2009, to February 25th, 2011. An application of our independence testing procedure to these series (without considering generalized error models) reveals, as expected, clear dependence between the series, the six statistics $\cW_n$, $\cF_n$, $\cH_n$, $\cH_{S,n}$, $\cH_{G,n}$ and $\cH_{E,n}$ rejecting the null hypothesis of independence with $P$-values nearly zero. The dependogram displayed in the left panel of Figure \ref{dep0}  shows significant dependence for $(X_t,X_t, Z_t)$, $(X_t,X_t, Z_{t-1})$ and
$(X_t,X_{t-2}, Z_t)$, where $X_t, X_t, Z_t$ represent respectively the daily returns of Apple, Intel, and Hewlett-Packard.
Next,  we tested the independence between generalized errors by fitting AR-Gaussian HMM model for each series, choosing the number of regimes as the first number where the $P$-value of the goodness-of-fit statistic is at least 5\%. However, since Intel has 14 observations with $0$ returns (2.6\%), its distribution is not continuous, and we considered a zero-inflated Gaussian ARXHMM, where the first regime having always value $0$; see, e.g., \cite{Nasri/Remillard/Thioub:2024}. For each time series, the retained models had three regimes. We retained a AR(1)-Gaussian HMM for the Apple, a zero-inflated AR(1)-Gaussian HMM for the Intel, while for the Hewlett-Packard series, we selected an AR(2)-Gaussian HMM. The estimated coefficients are
reported in Table \ref{fit1}. It is worth noting that the stationary distribution of the model for Intel is given $0.0253$, $0.2347$, and $0.7400$ respectively for regimes $1,2,3$.
\begin{table}[h!]
\caption{Estimated coefficients for the AR-Gaussian HMM model for Apple, Intel and Hewlett-Packard series. Note that for Intel, the first regime is the zero-inflated regime.}   \label{fit1}
\centering
\begin{tabular}{l|c|cc|c|c|ccc}
& & \multicolumn{2}{c|}{AR terms}& Constant &Sigma&\multicolumn{3}{c}{Transition probabilities}\\ \hline
Series& Regime j&  $\phi_{1,j}$ & $\phi_{2,j}$& $\theta_{1,j}$& $\sigma_j$ & $Q_{1,j}$ & $Q_{2,j}$& $Q_{3,j}$ \\ \hline
     &1 & -0.6232  &           &\;0.0095 & 0.0228 & 0.0693  &  0.2761  &  0.6546\\
Apple&2 &  \;0.7846  &           &-0.0069 & 0.0122 & 0.2398  &  0.1986  &  0.5615\\
     &3 & -0.0388  &           &\;0.0040 & 0.0100 & 0.4944  &  0.4944  &  0.0112\\  \hline
     &1 &   0     &           &   0     &    0    &   0      &  0.3718  & 0.6282  \\
Intel&2 &-0.1408  &           &-0.0008  &  0.0296 & 0.0420   &  0.9387  & 0.0193  \\
    & 3 &-0.0176  &           & 0.0010  &  0.0141 & 0.0209   &  0.0067  & 0.9724   \\ \hline
     &1 &  -0.0507   &   -0.0512   & -0.0040 & 0.0326   & 0.9603  &  0.0000  &   0.0397  \\
 HP  &2 &  -0.0403   & \;0.0010   &\;0.0036 & 0.0082   & 0.0000  &  0.9272  &   0.0728 \\
     &3 & \;0.0093   &   -0.0815   &\;0.0002 & 0.0158   & 0.0156  &  0.0415  &   0.9429  \\ \hline
\end{tabular}
\end{table}
Applying our independence testing procedures on the generalized errors of these models reveals that the generalized errors are not independent, all six statistics $\cW_n$, $\cF_n$, $\cH_n$, $\cH_{S,n}$, $\cH_{G,n}$ and $\cH_{E,n}$ rejecting the null hypothesis of independence with $P$-values near zero. The dependogram presented in the right  panel of Figure \ref{dep0} shows the only significant dependence is between the generalized errors of three models at the same time $t$; that is, there is no significant lagged dependence left, which makes more sense in this case, due to market efficiency.
Next, since all three stocks are part of the NASDAQ, we fitted AR-Gaussian HMM with the NASDAQ's return as a covariate for each series. The retained models were AR(1)X-Gaussian HMM with two regimes for Apple and Hewlett-Packard, and  a zero-inflated AR(1)-Gaussian HMM with 3 regimes for Intel. The estimated parameters are displayed in Table \ref{fit2}.
\begin{table}
\caption{Estimated AR-Gaussian HMMs with NASDAQ as a covariate, for Apple, Intel and HP daily returns.}   \label{fit2}
\centering
\begin{tabular}{l|c|c|cc|c|ccc}
& & AR & \multicolumn{2}{c|}{Covariates}&Sigma&\multicolumn{3}{c}{Transition probabilities}\\ \hline
Series& Regime j&  $\phi_{1,j}$ & $\theta_{1,j}$ & $\theta_{2,j}$& $\sigma_j$ & $Q_{1,j}$ & $Q_{2,j}$& $Q_{3,j}$ \\ \hline
Apple&1&   0.1978 &  0.0035   &  0.8650   &  0.0178  & 0.6735  & 0.3265 & \\
     &2&  -0.0077 & -0.0000   &  1.0288   &  0.0063  & 0.1965  & 0.8035 & \\ \hline
&     1&     0      &      0     &     0    &     0    &     0    & 1.0000 &  0.0000 \\
Intel &2& -0.0039    &   0.0016   &  1.0783  &  0.0150  &  0.0303  & 0.7068 &  0.2629    \\\
 &    3& -0.0104    &  -0.0013   &  1.0219  &  0.0076  &  0.0238  & 0.1700 &  0.8062      \\   \hline
 HP & 1&    0.0931    &  -0.0019   &   0.8825  &   0.0208  & 0.9738 & 0.0262  &   \\
    & 2&   -0.0502    &   0.0001   &   0.9354  &   0.0080  & 0.0128 & 0.9872  &     \\ \hline
\end{tabular}
\end{table}
We then applied our independence tests on the estimated generalized errors of the above models to check if NASDAQ's returns account for all the dependence between these series.
The respective $P$-values for the statistics $\cW_n$, $\cF_n$, $\cH_n$ $\cH_{S,n}$, $\cH_{G,n}$ and $\cH_{E,n}$ are $93.8\%$, $93.0\%$, $94.2\%$, $94.6\%$, $84.0\%$ and $6.0\%$ showing that the independence hypothesis is not rejected. This allows us to conclude that including the NASDAQ as
a covariate for the three series helped to account for the dependence. In other words, as described in the Introduction, the three series are conditionally independent given the past and Nasdaq. Looking also at the dependogram shown in Figure \ref{dep123}, we notice that there is no significant dependence left at any of the lags we considered.

\subsection{Application to the Pittsburgh crime data}\label{ssec:crime}

\begin{figure}[h!]
\centering
\includegraphics[scale = 0.5]{"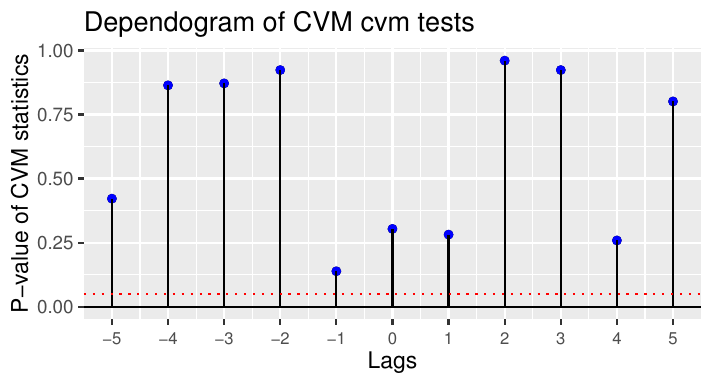"}
\includegraphics[scale = 0.5]{"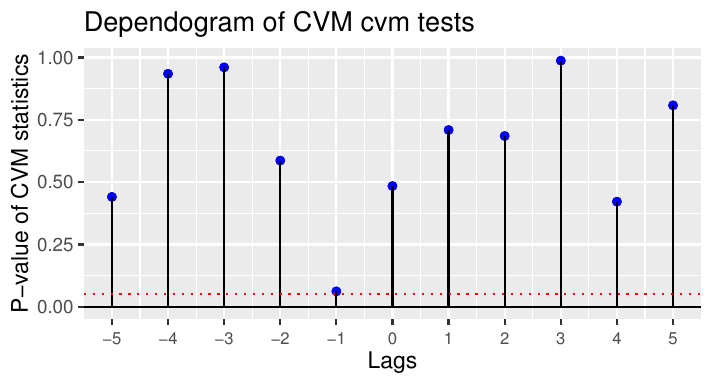"}
\caption{Dependograms for ROB and CMIS without fitting (left panel) and with INGARCH fitting (right panel).} \label{fig:pitts1}
\end{figure}

\begin{figure}[h!]
\centering
\includegraphics[scale = 0.5]{"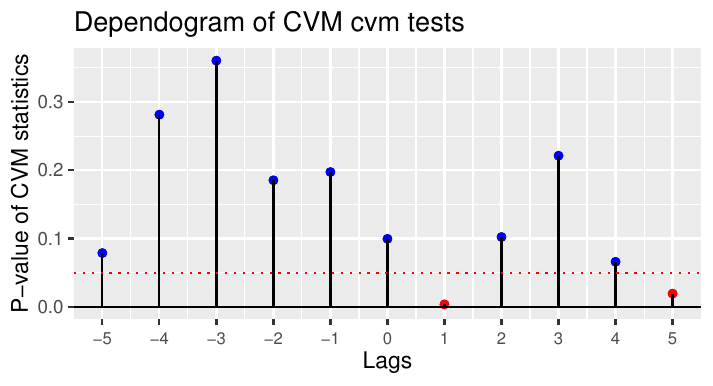"}
\includegraphics[scale = 0.5]{"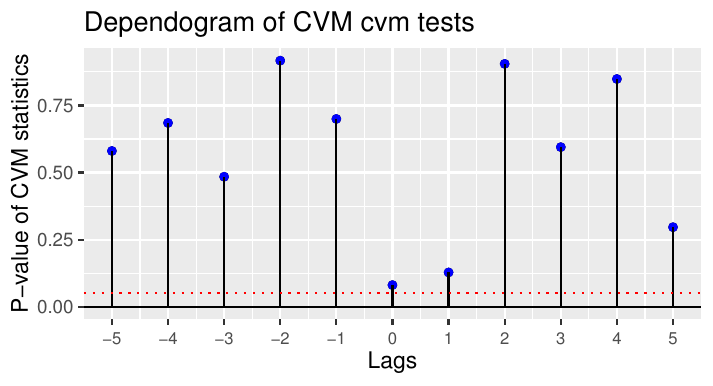"}
\caption{Dependograms for ROB and MVT without fitting (left panel) and with INGARCH fitting (right panel).} \label{fig:pitts2}
\end{figure}

As an example of analysis of discrete time series, we applied our methodology to Pittsburgh crime data for the period 1990--1999. These data were also considered by \cite{Cui/Zhu:2018}. For the sake of comparison,  we restricted the analysis to the same census tract $101$ studied in \cite{Cui/Zhu:2018}, and we first used the same two series, namely Robbery (ROB) and Crime mischief (CMIS).
The correlation between the two series is $-0.13$. Our statistical procedures were applied to test for independence between the two time series (without fitting generalized error models), and none of our statistics rejected the independence hypothesis. The $P$-values for $\mathcal{W}_n, \mathcal{F}_n$, $\mathcal{H}_n$ $\mathcal{H}_{S,n}$, $\mathcal{H}_{G,n}$ and $\mathcal{H}_{E,n}$ are respectively $65.2\%$, $66.4\%$, $69.5\%$, $90.4\%$, $91.6\%$ and $88.2\%$; see also the dependogram given in the left panel of Figure \ref{fig:pitts1}. Next, since there is serial dependence in each series, we fitted individual INGARCH(1,1) to each series, where the INGARCH(p,q) \citep{Ferland/Latour/Oraichi:2006} is an integer-valued process
$\{X_t\}$ such that the distribution of $X_t$ given the past sigma-field $\mathcal{F}_{t-1}$ is a Poisson with mean $\disp \lambda_t=\omega+\sum_{i=1}^q\alpha_i X_{t-i}+\sum_{j=1}^p\beta_j \lambda_{t-j}$ where $\omega >0$, $\alpha_1.\ldots,\alpha_q,\beta_1,\ldots,\beta_p$ are non-negative, with $\disp \sum_{i=1}^q \alpha_i +\sum_{j=1}^p\beta_j <1$. The estimated parameters were $\hat{\omega}=0.1187$, $\hat{\alpha}_1=0.0575$ and $\hat{\beta}_1=0.8849$ for the ROB series and $\hat{\omega}_0=2.654$, $\hat{\alpha}_1=0.307$ and $\hat{\beta}_1=0.252$ for the CMIS series.  We then tested if there was dependence left between the generalized errors of these models. The dependogram given in the right panel of Figure \ref{fig:pitts1} shows that no significant dependence exists for lags ranging from $-5,\ldots, 5$. The $P$-values of the statistics $\mathcal{W}_n, \mathcal{F}_n$, $\mathcal{H}_n$ $\mathcal{H}_{S,n}$, $\mathcal{H}_{G,n}$ and $\mathcal{H}_{E,n}$ are respectively $92.0\%$, $91.5\%$, $91.2\%$, $96.0\%$, $96.6\%$ and $93.2\%$. Therefore, we conclude that no significant dependence is left after fitting the individual INGARCH(1,1) to each series.
To demonstrate the use of more complex autoregressive Hidden Markov Models given in Definition \ref{ARX}, we fitted AR$(p)$-Poisson HMM to each time series. The number of regimes and the number of lags were identified using goodness-of-fit tests and BIC criteria. The retained models were AR(2) Poisson for the Robbery series and an AR(1) Poisson HMM with two regimes for the Crime mischief series. The estimated parameters for the ROB model were $\theta=1.6021$, $\phi_{1}=0$,  and  $\phi_{2}=0.2472$, while for CMIS the estimated parameters are   $\theta_{1}=6.2942$, $\phi_{1,1}=0.4511$, $\theta_{2}=2.0347$, $\phi_{1,2}= 0.3308$,  and
$\disp
\hat{Q}=\left(\begin{array}{ll}0.3432  &  0.6568 \\  0.4573  &  0.5427 \end{array} \right)$.
We then tested for any dependence left between the generalized errors of the two time series, and found that the $P$-values of the statistics $\mathcal{W}_n, \mathcal{F}_n$, $\mathcal{H}_n$ $\mathcal{H}_{S,n}$, $\mathcal{H}_{G,n}$ and $\mathcal{H}_{E,n}$ are respectively $79.6\%$, $81.7\%$, $81.1\%$, $87.0\%$, $84.4\%$ and $83.7\%$.  Therefore, in concordance with the above analysis,  we again conclude that no significant dependence is left between the generalized errors of the two time series, so that the two series are conditionally independent given the past.
We also considered the motor vehicle theft (MVT) series and examined the dependence between Robbery and Motor vehicle theft. We noticed that the correlation between ROB and MVT is $0.21$ and that when we applied our test procedures to these two series, the $P$-values of the statistics $\mathcal{W}_n, \mathcal{F}_n$, $\mathcal{H}_n$ $\mathcal{H}_{S,n}$, $\mathcal{H}_{G,n}$ and $\mathcal{H}_{E,n}$ are respectively $5.00\times 10^{-10}$, $8.81\times 10^{-10}$, $0.0594$, $7.26\times 10^{-5}$, $0.00118$ and $0.00106$. Clearly, all statistics, but $\cH_n$, reject the independence hypothesis. The dependogram displayed on the left side of Figure \ref{fig:pitts2}, reveals significant dependence at lags $1$ and $5$.  Fitting INGARCH(1,2) to the MVT series yielded the following estimated parameters $\hat{\omega}_0=0.696$, $\hat{\alpha}_1=0.411$, $\hat{\alpha}_2=0.212$ and $\hat{\beta}_1=0.243$. We then used our proposed test statistics to check if there is any dependence left between the generalized errors of these two series. None of our test statistics rejected the independence hypothesis and the $P$-values for $\mathcal{W}_n, \mathcal{F}_n$, $\mathcal{H}_n$ $\mathcal{H}_{S,n}$, $\mathcal{H}_{G,n}$ and $\mathcal{H}_{E,n}$ were respectively $71.4\%$, $75.2\%$, $74.2\%$, $64.7\%$, $44.6\%$ and $5.27\%$; see the right-side of Figure \ref{fig:pitts2} for illustration. This illustrates that the dependence between these two series is explained by the dynamic models fitted to each series. The result might sound surprising, as one expects other factors to explain the interdependence between these series. We argue that the conditional independence observed here is the result of the fact that these series are for the same small geographical location where factors that affect such dependence might have  constant values.

\begin{appendix}

\section{Auxiliary results}
\subsection{Expression for covariance}\label{app:cov}

Suppose that $X\sim G$ and let $\cA_G$ be the set of atoms of $X$ and define $g(y)=\Delta G(y)$. Then, for any $u,v\in [0,1]$
\begin{eqnarray*}
\chi_G(u,v)&=&E\left\{J_G(X,u) J_G(X,v)\right\}   -uv = G\circ G^{-1}(u\wedge v)-\sum_{y\le G^{-1}(u\wedge v)}g(y)  -uv \nonumber\\
&& \quad +\sum_{y\in\cA_G}g(y)D\left\{\frac{u-G(y-)}{g(y)}\right\} D\left\{\frac{v-G(y-)}{g(y)}\right\} = u\wedge v -uv - \chi_{0,G}(u,v),\label{eq:cov-Igen}
\end{eqnarray*}
where
$\disp  \chi_{0,G}(u,v)= \frac{\{u\wedge v -G(x-)\}\{G(x)-u\vee v\}}{g(x)}\I\left\{G^{-1}(u)=G^{-1}(v) \in \cA_G\right\}$,
with $x=G^{-1}(u)$.
Expressions for $\chi_G(u,v)$ and $\chi_G(u,v)$ agree with the ones in \cite{Kheifets/Velasco:2017} if $G$ is  a discrete cdf.

\subsection{Expression for $d_{G,F}$}\label{app:dgf}

Here, we extend the function of $d_{G,F}$  in \cite{Kheifets/Velasco:2017} 
to arbitrary distributions.
\begin{prop}\label{prop:dGF}
Assume that $F$ and $G$ have the same set of atoms, denoted by $\cA$ and set $d_{G,F}(u)=E_G\{J_F(Y,u)\}-u$. Then, if $F^{-1}(u)\notin \cA$, $d_{G,F}(u)=G$$\circ F^{-1}(u)-F\circ F^{-1}(u)=G\circ F^{-1}(u)-u $. If  $F^{-1}(u)\in \cA$, then
$$
    d_{G,F}(u)= G\circ F^{-1}(u)-F\circ F^{-1}(u) -\left[ \frac{F\circ F^{-1}(u)-u}{f\circ F^{-1}(u)}\right] \left[g\circ F^{-1}(u)-f\circ F^{-1}(u)\right].
    $$
Furthermore,
\begin{equation}\label{eq:d(G,F)}
    \sup_{u\in [0,1]} |d_{G,F}(u)|\le 3 \sup_{y\in\dR}|G(y)-F(y)| = 3\|G-F\|_\infty.
\end{equation}
\end{prop}
\noindent \textit{Proof.}
    Further set
$f(y)=\Delta F(y)$ and $g(y)=\Delta G(y)$. It then follows that if $F^{-1}(u)\in \cA$, then
\begin{eqnarray*}
    d_{G,F}(u)&=& \sum_{y\in\cA}  g(y)  D\left\{\frac{u-F(y-)}{f(y)}\right\}  +E\left[\dI \{Y\notin\cA\}\dI\left\{Y\le F^{-1}(u)\right\}\right]-u\\
    &=& g\circ F^{-1}(u) \left[ \frac{u-F\{F^{-1}(u)-\}}{f\circ F^{-1}(u)}  \right]+\sum_{y<F^{-1}(u)}g(y)
     +G\circ F^{-1}(u)-\sum_{y\le F^{-1}(u)}g(y) -u\\
    &=& G\circ F^{-1}(u)-F\circ F^{-1}(u) -\left[ \frac{F\circ F^{-1}(u)-u}{f\circ F^{-1}(u)}\right] \left[g\circ F^{-1}(u)-f\circ F^{-1}(u)\right],
\end{eqnarray*}
while if $F^{-1}(u)\notin \cA$, then $d_{G,F}(u)=G\circ F^{-1}(u)-F\circ F^{-1}(u)=G\circ F^{-1}(u)-u $, proving \eqref{eq:d(G,F)}.  \qed

\subsection{Bound for $E_G\left[\left\{J_G(Y,u)-J_F(Y,u)\right\}^2 \right]$}\label{app:bounds}
The following proposition extends and improve the upper bound for $E_G\left[\left\{J_G(Y,u)-J_F(Y,u)\right\}^2 \right]$ in \cite[Lemma A(iv)]{Kheifets/Velasco:2017}.
\begin{prop}\label{prop:boundJGF}
Assume that $F$ and $G$ have the same set $\cA$ of atoms. Then,
$ H(u) = E_G\left[\left\{J_G(Y,u)-J_F(Y,u)\right\}^2 \right] \le 3\sup_{y}|G(y)-F(y)| = 3\|G-F\|_\infty$.
\end{prop}
\noindent \textit{Proof.}
 As noted in \cite{Kheifets/Velasco:2017}, since   $|J_G(Y,u)-J_F(Y,u)|$ is bounded by $1$, it suffices to show that
 $E_G\left\{|J_G(Y,u)-J_F(Y,u)|\right\} \le 3\sup_{y}|G(y)-F(y)|$. To prove this, we consider four cases.\\
 \textbf{Case 1}: $G^{-1}(u)\notin \cA$. In this case, $G\circ G^{-1}(u)=u =F\circ F^{-1}(u)$, and
\begin{eqnarray*}
 H(u)&=&  E\left[\dI\{Y\notin \cA\}|\dI\{Y\le G^{-1}(u)\}-\dI\{Y\le F^{-1}(u)\}\right]  +\sum_{y}g(y) \left|D\left\{\frac{u-G(y-)}{g(y)}\right\}-D\left\{\frac{u-F(y-)}{f(y)}\right\}\right|\\
 &=& \left|G\circ F^{-1}(u)-u\right|-\sum_{G^{-1}(u)\wedge F^{-1}(u) < y< G^{-1}(u)\vee F^{-1}(u)}g(y)
 +\sum_{G^{-1}(u)\wedge F^{-1}(u) < y< G^{-1}(u)\vee F^{-1}(u)}g(y)  \\
 &=&  \left|G\circ F^{-1}(u)-F\circ F^{-1}(u)\right|\le \|G-F\|_\infty.
\end{eqnarray*}

\textbf{Case 2}: $y_0=G^{-1}(u) = F^{-1}(u) \in \cA$. In this case, since $\frac{F(y_0)-u}{f(y_0)}\in [0,1]$,
\begin{eqnarray*}
 H(u)&=&  E\left[\dI\{Y\notin \cA\}|\dI\{Y\le y_0\}-\dI\{Y\le y_0\}\right] +\sum_{y}g(y) \left|D\left\{\frac{u-G(y-)}{g(y)}\right\}-D\left\{\frac{u-F(y-)}{f(y)}\right\}\right|\\
 &=& g(y_0) \left|D\left\{\frac{u-G(y_0-)}{g(y_0)}\right\}-D\left\{\frac{u-F(y_0-)}{f(y_0)}\right\}\right|  +\sum_{y\neq y_0}g(y) \left|D\left\{\frac{u-G(y-)}{g(y)}\right\}-D\left\{\frac{u-F(y-)}{f(y)}\right\}\right|\\
  &=& g(y_0) \left|\frac{G(y_0)-u}{g(y_0)}-\frac{F(y_0)-u}{f(y_0)}\right|
  = \left|G(y_0)-u -\{ g(y_0)-f(y_0)\}\left\{\frac{F(y_0)-u}{f(y_0)}\right\} -F(y_0)+u \right| \\
  &\le & |G(y_0)-F(y_0)|+|g(y_0)-f(y_0)|\le 3 \|G-F\|_\infty.
\end{eqnarray*}

\textbf{Case 3}: $y_0 = G^{-1}(u) > F^{-1}(u) =y_1 \in \cA$. Then,
\begin{multline*}
 H(u)=  E\left[\dI\{Y\notin \cA\}|\dI\{Y\le y_0\}-\dI\{Y\le y_1\}\right] +\sum_{y}g(y) \left|D\left\{\frac{u-G(y-)}{g(y)}\right\}-D\left\{\frac{u-F(y-)}{f(y)}\right\}\right|\\
 = G(y_0)-G(y_1) -\sum_{y_1<y \le y_0}g(y) + g(y_0) \left|D\left\{\frac{u-G(y_0-)}{g(y_0)}\right\}-D\left\{\frac{u-F(y_0-)}{f(y_0)}\right\}\right|  \\
  + g(y_1) \left|D\left\{\frac{u-G(y_1-)}{g(y_1)}\right\}-D\left\{\frac{u-F(y_1-)}{f(y_1)}\right\}\right|
  +\sum_{y_1<y<y_0 }g(y) \left|D\left\{\frac{u-G(y-)}{g(y)}\right\}-D\left\{\frac{u-F(y-)}{f(y)}\right\}\right|\\
 =G(y_0)-G(y_1) -\sum_{y_1<y \le y_0}  g(y) + u-G(y_0-) + g(y_1)\left\{\frac{F(y_1)-u}{f(y_1)}\right\}  +\sum_{y_1<y<y_0 }g(y)\\
=  F(y_1)-G(y_1)  +\{ g(y_1)-f(y_1)\}\left\{\frac{F(y_1)-u}{f(y_1)}\right\}
\le|G(y_1)-F(y_1)|+|g(y_1)-f(y_1)|\le 3 \|G-F\|_\infty.
\end{multline*}

\textbf{Case 4}: $y_0 = G^{-1}(u) < F^{-1}(u) =y_1\in \cA$. Then, the proof is similar to Case 3.
\qed

\begin{prop}\label{prop:boundV2} For any interval $B(\epsilon)\subset [0,1]$ of length $\epsilon^2$,
$\disp E_G\left[\sup_{u,v\in B(\epsilon)} |J_G(Y,u)-u-J_G(Y,v)+v|^2 \right] \le 4\epsilon^2$.
\end{prop}

\noindent \textit{Proof.}
Without loss of generality, assume that $u<v$ and that $B(\epsilon)=[a,b]\subset [0,1]$ such that $b-a=\epsilon^2$. One first notices that
$|J_G(Y,u)-u-J_G(Y,v)+v|\le 1$ and that $|J_G(Y,u)-u-J_G(Y,v)+v|=|u-v|\le \epsilon^2$ if  $G(Y) < a$ or $G(Y)>b$. Moreover, if $\Delta G(G^{-1}(a))>0$, then
for $Y=G^{-1}(a)$  and  $u,v \in [a,b]$, one has
  $\disp |J_G(Y,u)-u-J_G(Y,v)+v| =|u-v|\frac{1-\Delta G(G^{-1}(a))}{\Delta G(G^{-1}(a))}$
  when $G(G^{-1}(a)^-)<u\le v\le G(G^{-1}(a))$, and
  \begin{eqnarray*}
      |J_G(Y,u)-u-J_G(Y,v)+v| &=& \left|\frac{u-G(G^{-1}(a))}{\Delta G(G^{-1}(a))}-u+v\right|
          \le |u-v|\frac{\max\{1-\Delta G(G^{-1}(a)),\Delta G(G^{-1}(a))\}}{\Delta G(G^{-1}(a))},
  \end{eqnarray*}
   when $G(G^{-1}(a)^-)<u\le  G(G^{-1}(a))<v$,  and
  $|J_G(Y,u)-u-J_G(Y,v)+v| =|u-v|$ when $G(G^{-1}(a))<u\le v$. Hence, in all cases when $Y=G^{-1}(a)$ we have $|J_G(Y,u)-u-J_G(Y,v)+v| \le \epsilon^2/{\Delta G(G^{-1}(a))}$ for all $u,v\in [a,b]$. Similar steps repeated for $\Delta G(G^{-1}(b))>0$ show that when $Y=G^{-1}(b)$ we have $|J_G(Y,u)-u-J_G(Y,v)+v| \le \epsilon^2/{\Delta G(G^{-1}(b))}$ for all $u,v\in [a,b]$.
Therefore, one easily verifies that
\begin{multline*} E_G\left[\sup_{u,v\in B(\epsilon)} |J_G(Y,u)-u-J_G(Y,v)+v|^2 \right]\\
\le \epsilon^2 P\{Y< G^{-1}(a)\}  +\epsilon^2 P\{Y > G^{-1}(b)\}  +P\{G^{-1}(a)< Y < G^{-1}(b)\} \\
 +\epsilon^2/\Delta G(G^{-1}(a))P\{Y=G^{-1}(a)\}  +\epsilon^2/\Delta G(G^{-1}(b))P\{Y=G^{-1}(b)\}
 \le 4\epsilon^2,
\end{multline*}
completing the proof. \qed
\section{Proofs of the main results}\label{app:proofs}

\subsection{Proof of Theorem \ref{thm:mainK}}
The proof is very similar to the proof of Theorem 1 in \cite{Nasri/Remillard:2019}, who studied the asymptotic of empirical processes
$\disp
\dK_n(\by)=n^{-\frac{1}{2}}\sum_{t=1}^n \left[\dI\{\be_{n,t}\le \by\}-K(\by)\right]$, where $K$ is the joint cdf of $\bve_t$,
and $\mathbb{F}_{nj}(x)=\sqrt{n}\{F_{nj}(x)-F_j(x)\}$ for continuous generalized error models.  Here, under the null hypothesis of independence, $K(\by) = \prod_{j=1}^d F_j(y_j)$, and the $\bve_t$ are iid.  In our general case, taking $\bfl\equiv 0$ for simplicity, we have that
\begin{eqnarray*}
  \dK_n(\by) - \alpha_n(\by) &=&  n^{-\frac{1}{2}}\sum_{A\subset \cS_d\setminus\emptyset} \sum_{t=1}^n \prod_{j\in A}\left[\dI\{e_{n,jt}\le y_j\}-\dI\{\ve_{jt}\le y_j\}\right]\prod_{j\not \in A}\dI\{\ve_{jt}\le y_j\}\\
  &=& n^{-\frac{1}{2}}\sum_{j=1}^d \sum_{t=1}^n \left[\dI\{e_{n,jt}\le y_j\}-\dI\{\ve_{jt}\le y_j\}\right]\prod_{k\neq j}\dI\{\ve_{jt}\le y_j\}+o_P(1),
\end{eqnarray*}
using Assumptions \ref{A0}--\ref{A6}. Furthermore,  using \cite{Ghoudi/Remillard:2004} and   \cite{Nasri/Remillard:2019},
one gets that for any $j\in\setd$,
\begin{multline*}
    n^{-\frac{1}{2}}\sum_{t=1}^n \left[\dI\{e_{n,jt}\le y_j\}-\dI\{\ve_{jt}\le y_j\}\right]\prod_{k\neq j}\dI\{\ve_{jt}\le y_j\}
=f_j(y_j) \left\{\prod_{k\neq j}F_k(y_k)\right\} \frac{1}{n} \sum_{t=1}^n \bgamma_{jt}\circ F_j(y_j)^\top \bTheta_n+o_P(1),
\end{multline*}
so, under Assumptions \ref{A0}--\ref{A6},   $\dK_n$ converges in $D([-\infty,\infty]^d)$ to $\dK$,
where $$\dK(\by)=\beta\circ K(\by)-\sum_{j=1}^d f_j(y_j)\left\{\prod_{k\neq j}F_k(y_k)\right\}\bGamma_j(y_j)^\top \bTheta.
$$
As a by-product, the processes $\mathbb{F}_{n,j}$, $j\in\setd$,  converge jointly in $D([-\infty,\infty])$ to
$\mathbb{F}_{j}$, with $\mathbb{F}_{j}(x_j)=\beta^{(j)}\circ F_j(x_j)-f_j(x_j)\Gamma_j(x_j)^\top \bTheta$. \qed

\subsection{Proof of Corollary \ref{cor:mainK}}
For any finite set $L \subset \dZ^d$, it is well known that the processes $\{\alpha_{n,\bfl}; \bfl \in L\}$ converge jointly in
$D([0,1]^d)$ to continuous centered Gaussian processes $\{\alpha_\bfl;\bfl\in L\}$ having the same distribution as
$\alpha$, which is a $d$-dimensional Brownian bridge. See, e.g.,  \cite{Bickel/Wishura:1971}.
One easily notices that the process $\mathcal{K}_{n,\bfl}(\bx)=\dK_{n,\bfl}(\bx)-n^{\frac{1}{2}}\left[\prod_{j=1}^d F_{nj}(x_j)-\prod_{j=1}^d F_{j}(x_j)\right]$.
By the above result, $\dK_{n,\bfl}(\bx)$ converges weakly to $\dK_{\bfl}(\bx)=\alpha_{\bfl}(K(\bx))-\nabla K(\bx)\bGamma(\bx)\bTheta$. Using the multinomial formula and the convergence of the processes $\dF_{nj}
$,
one has
\begin{eqnarray*}
 n^{\frac{1}{2}}\left\{\prod_{j=1}^d F_{nj}(y_j) - \prod_{j=1}^d F_j(y_j)\right\}
& = &
 n^{\frac{1}{2}}\left[\prod_{j=1}^d \left\{F_{j}(y_j)+ n^{-\frac{1}{2}}\dF_{nj}(y_j)\right\} - \prod_{j=1}^d F_j(y_j)\right] \\
&=&
 \sum_{j=1}^d \mathbb{F}_{nj}(y_j)\left\{\prod_{k\neq j} F_k(y_k)\right\} +O_P(n^{-\frac{1}{2}}),
\end{eqnarray*}
which converges in $D([-\infty,\infty]^d)$ to
$$
\sum_{j=1}^d \dF_j(y_j)\left\{\prod_{k\neq j}^d F_k(y_k)\right\}=\sum_{j=1}^d \left\{\prod_{k\neq j}^d F_k(y_k)\right\}\left[\beta^{(j)}
 \{F_j(y_j)\}- f_j(y_j)\bGamma_j(y_j)^\top \bTheta\right].
$$
Combining the above results shows that $\mathcal{K}_{n,\bfl}$ converges weakly to $\mathcal{K}_{\bfl}=\dC_\bfl\{K$.
\qed

\subsection{Proof of Corollary \ref{cor:maincop}}
The convergence of $\dC_{n,\bfl}$ follows from the convergence of $\dK_{n,\bfl}$ in Theorem \ref{thm:mainK} together with Proposition A.1 in \cite{Genest/Ghoudi/Remillard:2007}. \qed

\subsection{Proof of Theorem~\ref{thm:cop}}

Using Theorem~\ref{thm:mainK} and Corollary \ref{cor:maincop}, the proof follows identical steps to those of Theorem ~2 in \cite{Duchesne/Ghoudi/Remillard:2012}. \qed

\subsection{Proof of Theorem \ref{thm:khf}} To simplify the proof, assume that $\bfl\equiv 0$ and set $G_{jt}=G_{\btheta_0,jt}$.
Before starting the proof, the following lemma summarizes some needed results.
\begin{lem}\label{lemres1} Let
$\disp V_t(\bu)=\prod_{j=1}^d J_{G_{jt}}(X_{jt},u_j) - \prod_{j=1}^d u_j$. Then,
\begin{eqnarray*}
|V_t(\bu)-V_t(\bv)|
& \le &  \sum_{j=1}^d |J_{G_{jt}}(X_{jt},u_j) -  J_{G_{jt}}(X_{jt},v_j) | +\sum_{j=1}^d |u_j-  v_j| \\
& \le &  \sum_{j=1}^d |J_{G_{jt}}(X_{jt},u_j) -  u_j-J_{G_{jt}}(X_{jt},v_j) +  v_j| +2\sum_{j=1}^d |u_j-  v_j| .
\end{eqnarray*}
\end{lem}
\noindent \textit{Proof.}
The proof is straightforward since all terms are bounded by $1$.
\qed

 Now, to prove the Theorem, note that under $H_0$, the sequence $\disp V_t(\bu)$
 is a martingale difference sequence with respect to $\mathcal{F}_t$, using \eqref{eq:sklarmaltese}. Next,  observe that based on the results in Appendix \ref{app:cov},
$$
  \kappa(\bu,\bv) =
  \lim_{n\to\infty}
  \frac{1}{n}\sum_{t=1}^n  \prod_{j=1}^d E\left\{J_{G_{j,t}}(X_{j,t},u_j)  J_{G_{j,t}}(X_{j,t},v_j)|\cF_{t-1}  \right\} - \prod_{j=1}^d (u_j v_j),
$$
and
$$
\kappa_{A}(\bu,\bv) = \lim_{n\to\infty}
\frac{1}{n}\sum_{t=1}^n \prod_{j\in A}\left[ E\left\{
J_{G_{j,t}}(X_{j,t},u_j)
J_{G_{j,t}}(X_{j,tj},v_j) |\cF_{t-1}\right\}-u_j v_j\right],
$$
coincide with the expressions in Assumption (A7').
In order to establish the convergence of $\tilde\dK_n$ and $\tilde\dK_{n,A}$, one needs to first study the convergence of $\tilde\beta_n$  and $\tilde\beta_{n,A}$.
The convergence of $\tilde\beta_n$ follows  by showing that the conditions of Theorem A in  \cite{Kheifets/Velasco:2017} hold.
In fact, Assumption (A7') is equivalent to their condition (N1), while condition (N2) translates to $ \frac{1}{n}\sum_{t=1}^nE[\|V_t\|^2\dI\{\|V_t\|>\epsilon \sqrt{n}\}|\mathcal{F}_{t-1}]$ converges in probability to zero for every $\epsilon >0$, as $n\to\infty$. The latter condition is trivially met  since $\|V_t\|\le 2$.
 To check their entropy condition (N3), we choose a partition of $[0,1]^d$ obtained by partitioning each dimension into
  $0=u_0<u_1\ldots < u_{N_\epsilon^0}=1$ such that $u_{i+1}-u_i\le \epsilon^{2}$ for all $i=1,\ldots,N_{\epsilon}^0-1$. Note that $N_\epsilon^0=\lfloor\epsilon^{-2}\rfloor+1$ for $\epsilon <1$ and $N_\epsilon^0=1$ for $\epsilon \ge 1$. The number of elements in the generated partition of $[0,1]^d$ is $N_\epsilon=(N_\epsilon^0)^d$, which satisfies $\int_0^\infty\sqrt{\log(N_\epsilon)}d\epsilon=\int_0^1\sqrt{\log(N_\epsilon)}d\epsilon <\infty$. It remains to check that
  \begin{equation}\label{W2}\sup_{\epsilon\in (0,1)\cap\mathbb{Q}}\frac{1}{n\epsilon^2}\max_{1\le k\le N_\epsilon}{\sum_{t=1}^n E\left[\sup_{\bu,\bv\in \mathcal{B}_k(\epsilon)}(V_t(\bu)-V_t(\bv))^2 \big|\mathcal{F}_t\right]}=O_p(1).
  \end{equation}
 From Lemma \ref{lemres1} and the fact that $|V_t(\cdot)|\le 1$,  one gets that
 $$
 |V_t(\bu)-V_t(\bv)|^2 \le 2  \sum_{j=1}^d |J_{G_{jt}}(X_{jt},u_j) -  u_j-J_{G_{jt}}(X_{jt},v_j) +  v_j| +4 \sum_{j=1}^d |u_j-  v_j|.
 $$
 Using Proposition \ref{prop:boundV2} in Appendix \ref{app:bounds}, one deduces that
 $\disp  E\left[\sup_{\bu,\bv\in \mathcal{B}_k(\epsilon)}\{V_t(\bu)-V_t(\bv)\}^2\big|\mathcal{F}_t \right] \le 12 d \epsilon^2$,
 which implies that \eqref{W2} holds and yields the convergence of $\tilde\beta_{n}$. Also, the convergence of $\tilde\beta_{n,A}$ holds since it can be written as a finite sum of the processes
 $\tilde\beta_n$ evaluated at different points $\bu$. More precisely,  $\tilde\beta_{n,A}=\sum_{B\subset A} \tilde\beta_{n}(\bu_B)\prod_{k\in A\setminus B}u_k$ where $\bu_B$ is $d$-dimensional vector whose $j$th component ${\bu_B}_j=u_j$ if $j\in B$ and ${\bu_B}_j=1$ if $j\notin B$.
Now, to prove the convergence of $\tilde\dK_n$, note that $\tilde\dK_n(\bu)=\tilde\beta_n(\bu)+\tilde\delta_n(\bu)$, where
$$
\tilde\delta_n(\bu)=n^{-\frac{1}{2}}\sum_{t=1}^n \left[ \prod_{j=1}^d J_{G_{\btheta_n,jt}}(X_{jt},u_j) - \prod_{j=1}^d J_{G_{jt}}(X_{jt},u_j) \right].
$$
Using the multinomial formula, one  gets
$\disp
\tilde\delta_n(\bu)=\sum_{j=1}^d \tilde\delta_{n,1j}(\bu)+\sum_{A\subset \cS_d,|A|>1} \tilde\delta_{n,2,A}(\bu)$,
where
$$
\tilde\delta_{n,1j}(\bu)=n^{-\frac{1}{2}}\sum_{t=1}^n \left\{J_{G_{\btheta_n,jt}}(X_{jt},u_j) -  J_{G_{jt}}(X_{jt},u_j)\right\}\prod_{k\neq j}
J_{G_{kt}}(X_{kt},u_k)
$$
and
$$
\tilde\delta_{n,2,A}(\bu)=n^{-\frac{1}{2}}\sum_{t=1}^n \prod_{j\in A}\left\{J_{G_{\btheta_n,jt}}(X_{jt},u_j) -  J_{G_{jt}}(X_{jt},u_j)\right\}\prod_{k\notin A} J_{G_{kt}}(X_{kt},u_k).
$$
It will be shown that $\tilde\delta_{n,2,A}(\bu)=o_p(1)$ and that
$$
\sup_{\bu\in[0,1]^d}\left|\tilde\delta_{n,1,j}(\bu)-\bTheta_n^\top\frac{1}{n}\sum_{t=1}^n \bgamma_{jt}(u_j) \left\{\prod_{k\neq j} J_{G_{kt}}(X_{kt},u_k)\right\}\right|=o_p(1).
$$
To achieve this, decompose  $\tilde\delta_{n,1,j}$  into
$\disp \tilde\delta_{n,1,j}(\bu)= \tilde\delta_{n,1,1}^j(\bu,\btheta_n)+\tilde\delta_{n,1,2}^j(\bu,\btheta_n)$, where for any $\bbeta\in \mathbb{R}^p$,
$\disp
\tilde\delta_{n,1,1}^j(\bu,\bbeta)=n^{-\frac{1}{2}}\sum_{t=1}^n  \xi_{jt}(\bu,\bbeta)$,
with
$$
\xi_{jt}(\bu,\bbeta)=\left[J_{G_{\bbeta,jt}}(X_{jt},u_j) -  J_{G_{jt}}(X_{tj},u_j)-d_{G_{jt},G_{\bbeta,jt}}(u_j)\right]\prod_{k\neq j} J_{G_{kt}}(X_{kt},u_k)
$$
and
$\disp \tilde\delta_{n,1,2}^j(\bu,\bbeta)=n^{-\frac{1}{2}}\sum_{t=1}^n d_{G_{jt},G_{\bbeta,jt}}(u_j)\prod_{k\neq j} J_{G_{kt}}(X_{kt},u_k)$.
Next,
since $\bTheta_n $ is tight, we just need to consider $\bbeta$ such that $\|\bbeta-\btheta_0\|< n^{-\nu}$ for some $0<\nu<1/2$.
Proceeding as in \cite{Kheifets/Velasco:2017},  one verifies that $\tilde\delta_{n,1,1}^j(\bu,\bbeta)$ is a sum of martingale differences. Using Doob's and Rosenthal's inequalities with $p=4$, one gets that for each $\epsilon>0$,
$\disp
P\left\{\sup_{1\le m\le n} |\tilde\delta_{m,1,1}^j(\bu,\bbeta)|>\epsilon\right\}\le E\left\{|\tilde\delta_{n,1,1}^j(\bu,\bbeta)|^4\right\}/\epsilon^4
$ and
$\disp
E\left\{|\tilde\delta_{n,1,1}^j(\bu,\bbeta)|^4\right\}\le C\left[ E\left\{\frac{1}{n}\sum_{t=1}^n E(\xi_{jt}^2|\mathcal{F}_{t-1})\right\}^{2}+\frac{1}{n^2}\sum_{t=1}^n E|\xi_{jt}|^4\right]$.
Since $\xi_{jt}\le 2$, one sees that $\frac{1}{n^2}\sum_{t=1}^n E\left(\xi_{jt}\right)^4\le \frac{2^4}{n}\to 0$. Moreover, from \eqref{eq:d(G,F)} and Proposition \ref{prop:boundJGF}, one deduces that
\begin{eqnarray*}
E\left(\xi_{jt}^2|\mathcal{F}_{t-1}\right)&\le& 6\sup_{y}|G_{\bbeta,jt}(y)-G_{jt}(y)|+2\sup_{u\in [0,1]}|d_{G_{jt},G_{\bbeta,jt}}(u)|^2\\
&\le& 24\disp \sup_{y}|G_{\bbeta,jt}(y)-G_{jt}(y)| \le  24\|\bbeta-\btheta_0\|\sup_{y}\sup_{\btheta\in\mathcal{B}(\delta)}\|\dot{G}_{\btheta,jt}(y)\|.
\end{eqnarray*}
Hence,
$\disp
E\left[ \left\{\frac{1}{n}\sum_{t=1}^n E(\xi_t^2|\mathcal{F}_{t-1})\right\}^2\right] \le 576\|\bbeta-\btheta_0\|^2\frac{1}{n}\sum_{t=1}^n E\left[\sup_y \sup_{\btheta\in\mathcal{B}(\delta)}\|\dot{G}_{\btheta,jt}(y)\|^2\right]$.
By Assumption (A5'), the above is $O_P\left(n^{-2\nu}\right)$, and goes to zero as $n\to\infty$. This establishes the pointwise convergence in probability of $\tilde\delta_{n,1,1}^j$ to zero. To show that the convergence is uniform in $\bu$ and $\bbeta$, we use the monotonicity of $J_G$ and the continuity
of $d_{G,F}(u)$. Next, by Assumption (A2'),
$\disp
\sup_{\bu}\left|\tilde\delta_{n,1,2}^j(\bu,\btheta_n)-\bTheta_n^\top\frac{1}{n}\sum_{t=1}^n \bgamma_{jt}(u_j) \prod_{k\neq j} J_{G_{kt}}(X_{kt},u_k)\right| = o_p(1)$.
To complete the proof, it only remains to show that, uniformly in $\bu$, $\tilde\delta_{n,2,A}(\bu)$ converges to zero in probability. The proof follows similar steps as for $\tilde\delta_{n,1}$. In particular, we write
 we write $\tilde\delta_{n,2,A}(\bu, \btheta_n)=\tilde\delta_{n,2,A,1}(\bu, \btheta_n)+\tilde\delta_{n,2,A,2}(\bu, \btheta_n)$ where
$\disp
\tilde\delta_{n,2,A,1}(\bu, \bbeta)=n^{-\frac{1}{2}}\sum_{t=1}^n \sum_{B\subset A, B\neq \emptyset}\xi_{t,A,B}(\bu,\bbeta)$,
with
\begin{multline*}
\xi_{t,A,B}(\bu,\bbeta)=\prod_{j\in B}\left\{J_{G_{\bbeta,jt}}(u_j) -  J_{G_{jt}}(u_j)-d_{G_{jt},G_{\bbeta,jt}}(u_j)\right\}
\times\left\{\prod_{j\in A\setminus B}d_{G_{jt},G_{\bbeta,jt}}(u_j)\right\}\prod_{k\notin A} J_{G_{kt}}(X_{kt},u_k)
\end{multline*}
and
$\disp \tilde\delta_{n,2,A,2}(\bu, \bbeta)=n^{-\frac{1}{2}}\sum_{t=1}^n \prod_{j\in A}d_{G_{jt},G_{\bbeta,jt}}(u_j)\prod_{j\notin A} J_{G_{jt}}(u_j)$.
Mimicking the same arguments as in the proof of $\tilde\delta_{n,1,1}^j$, one easily shows that $\tilde\delta_{n,2,A,1}$ is a sequence of martingale differences which converges to zero in probability.
To complete the proof, note that $\disp \left|\tilde\delta_{n,2,A,2}(\bu, \btheta_n)\right|\le n^{-\frac{1}{2}}\sum_{t=1}^n \prod_{j\in A}\left|d_{G_{jt},G_{\btheta_n,jt}}(u_j)\right|\le \Upsilon_n $, where
$\disp \Upsilon_n =n^{-\frac{1}{2}}\sum_{t=1}^n |d_{G_{j_1,t},G_{\btheta_n,j_1,t}}(u_{j_1})||d_{G_{j_2,t},G_{\btheta_n,j_2},t}(u_{j_2})|$, where $j_1<j_2 \in A$.
Using Proposition \ref{prop:dGF} in Appendix \ref{app:dgf},
one sees that $\Upsilon_n \le \frac{9}{\sqrt{n}}\sum_{t=1}^n\|G_{\btheta_n,jt}- G_{\btheta_0,jt}\|^2 = o_P(1)$,
by Assumptions (A5') and \ref{A0}. Finally, the convergence of $\tilde\dK_{n,A}$ is similar to the convergence of $\tilde \beta_{n,A}$. \qed

\section{Examples of dependence measures when the margins are not continuous}\label{app:dep-measures}

Suppose $(U_1,U_2)\sim C$ and $(V_1,V_2)\sim \Pi$ are independent. Set $U_i^\star = F_i(X_i-)+V_i\Delta F_i(X_i)$, where $X_i=F_i^{-1}(U_i)$, $i\in\{1,2\}$. Then,
\begin{eqnarray*}
 \rho_{K_1,K_2,F_1,F_2}  &=&    {\rm cor}\left\{K_1^{-1}\left(U_1^\star\right),K_2^{-1}\left(U_2^\star\right)\right\} =
    \frac{1}{\sigma_1\sigma_2}\int_0^1 \int_0^1 \left\{K_1^{-1}(u_1) -\mu_1\right\} \left\{K_2^{-1}(u_2) -\mu_2\right\}dC^\maltese(u_1,u_2)\\
    &=& \frac{1}{\sigma_1\sigma_2}\int_0^1 \int_0^1 \left\{\cK_{1,G_1}\circ F_1^{-1}(u_1)-\mu_1\right\} \left\{\cK_{2,F_2}\circ F_2^{-1}(u_2) -\mu_2\right\}dC(u_1,u_2).
\end{eqnarray*}
If $F_2$ is continuous and $F_1$ is Bernoulli with parameter $1-p$, i.e., $F_1(0)=p$, then one gets
\begin{eqnarray*}
 \rho_{K_1,K_2} &=& \rho_{K_1,K_2,F_1,F_2}=\frac{1}{\sigma_1\sigma_2}
\left[ \frac{1}{p} \int_0^p  \left\{K_1^{-1}(s)-\mu_1\right\}ds\right] E\left[ \I\{U_1\le p\}  \left\{K_2^{-1}(U_2)-\mu_2\right\} \right],\\
&& \qquad +\frac{1}{\sigma_1\sigma_2}
\left[\frac{1}{1-p}\int_p^1 \left\{K_1^{-1}(s)-\mu_1\right\}ds\right] E\left[ \I\{U_1> p\}  \left\{K_2^{-1}(U_2)-\mu_2\right\} \right].
\end{eqnarray*}
Suppose $C$ is the Tent Map copula and set $H_j(s) = \int_0^s \left\{K_j^{-1}(u)-\mu_j\right\}du $, $j\in\{1,2\}$. Then,
$
\rho_{K_1,K_2} = \frac{1}{\sigma_1\sigma_2 } \frac{H_1(p) H_2(2p)}{2p(1-p)} $ if $0<p \le \frac{1}{2}$, and
$
\rho_{K_1,K_2} = -\frac{1}{\sigma_1\sigma_2 } \frac{H_1(p) H_2(2(1-p))}{2p(1-p)} $,  if $\frac{1}{2} \le p<1$.
Since $H_j(1)=0$ by hypothesis, it follows that $\rho_{K_1,K_2} = 0$ when $p=1/2$.
In particular, for Spearman's rho, $H(u) = -\frac{u(1-u)}{2}$, so if $p\le \frac{1}{2}$, $\rho_S = 6p\left(\frac{1}{2}-p\right)$ and if $p\ge \frac{1}{2}$, $\rho_S = -6(1-p)\left(p-\frac{1}{2}\right)$. For Savage's coefficient, $H(u) = u\log{u}$, so $\rho_E = \frac{p\log{p}\log{(2p)}}{1-p}$ if $0<p\le \frac{1}{2}$, and if $\frac{1}{2}\le p<1$,
$\rho_E = - \log{p}\log{(2(1-p))}$.

\input{conditions}

\end{appendix}
\bibliographystyle{apalike}

\end{document}

%% file: conditions.tex
\section{Examples of checking assumptions for time series }\label{app:ex}

\subsection{ARMA(p,q)}
For simplicity, consider an $ARMA(1,1)$ process $Y$, defined by $Y_t -\mu  -  \phi(Y_{t-1}-\mu) = \ve_t -\theta\ve_{t-1}$, with $\max(|\phi|,|\theta|)<1$.
Then, for $t\ge 1$, $\ve_t = Y_t-\mu   + (\theta-\phi)\sum_{k=1}^{t-1} \theta^{k-1}(Y_{t-k}-\mu)+\theta^t Z_0$, where $Z_0 = (\theta-\phi)\sum_{k=0}^\infty \theta^{k-1}(Y_{-k}-\mu)$.
As a result, if $\tilde Y$ is the time series defined by the previous equation with $\tilde Z_0=0$, then
$\left|\tilde Y_t-Y_t\right|\le |\theta|^t |Z_0$, which tends to $0$ exponentially fast. Next, if the cdf $G$ of $\ve_t$ has a bounded and continuous density, then
$\left| P\left(Y_t\le y|\cF_{t-1}\right)- P\left(\tilde Y_t\le y|\cF_{t-1}\right) \right| \le |\theta|^t |Z_0|$, and
$$
G_t(y) = P\left( Y_t\le y|\cF_{t-1}\right)  = G \left\{  y -\mu + (\theta-\phi) \sum_{k=1}^\infty \theta^{k-1} \left( Y_{t-k}-\mu\right)\right\}.
$$
As a result,
$\disp \partial_\mu G_t \left\{G_t^{-1}(u) \right\} =  - g\circ G^{-1}(u) \left( \frac{1-\phi}{1-\theta} \right)$, $\disp 
\partial_\phi G_t \left\{G_t^{-1}(u) \right\} =  - g\circ G^{-1}(u) \left\{\sum_{k=1}^\infty \theta^{k-1} \left( Y_{t-k}-\mu\right) \right\}$, and $\disp 
 \partial_\theta G_t \left\{G_t^{-1}(u) \right\} =  g\circ G^{-1}(u) \left\{\sum_{k=1}^\infty \theta^{k-2} \left(\theta+(k-1)(\theta-\phi) \right)\left( Y_{t-k}-\mu\right) \right\}$. 
Hence, $\bgamma_t(u)$ is continuous on $[0,1]$, stationary and ergodic, and $\bGamma(u) = - g\circ G^{-1}(u) \left( \frac{1-\phi}{1-\theta} \right) \be_1
$, where $\be_1 = (1,0,0)^\top $. For an ARMA(p,q) model with $\phi(z) = 1-\sum_{k=1}^p \phi_k z^k$ and $\theta (z) = 1-\sum_{k=1}^q \theta_k z^k$, where the zeros of $\phi(z)$ and $\theta(z)$ are outside the unit ball, then
$\bGamma(u) =  - g\circ G^{-1}(u)  \frac{\phi(1)}{\theta(1) } (1,0,\ldots,0)^\top \in \dR^{p+q+1}$.

\subsection{AR(p)-Poisson}
Another interesting example is an $AR(1)$-Poisson process, where the conditional distribution of $X_t$ given the past is Poisson, with mean $\lambda_{t-1} = \mu+\phi X_{t-1}$, $\mu>0$, and $0\le \beta<1$. Note that if $F_\lambda$ is the cdf of a Poisson distribution with mean $\lambda$, with density $f_\lambda$, then $F_\lambda(y)$ is strictly decreasing in $\lambda$, because $\partial_\lambda F_\lambda(y) =-f_\lambda(y)$. As a result, $F_\lambda^{-1}(u)$ is non-decreasing in $\lambda$, and if $U_t$ is an i.i.d. sequence of uniform variates on $(0,1)$, then, for $t\ge 1$, $Y_t = F_{\lambda_{t-1}}(U_t) \ge \tilde Y_t = F_{\tilde \lambda_{t-1}}(U_t)$, where $\tilde Y_0=0$, $\tilde \lambda_{t-1} = \mu+\phi\tilde Y_{t-1}$, and $\tilde \lambda_0=\mu \le \lambda_0$, $t\ge 1$. Next, $E\left(Y_t-\tilde Y_t\right) = \phi^t E(Y_0)$. As a result, by Borel-Cantelli's lemma, if $0< \phi<\alpha 1$, then
$P\left(Y_t-\tilde Y_t\ge \alpha^t \right)\le \left(\frac{\phi}{\alpha}\right)^t E(Y_0)$, so $P\left(Y_t-\tilde Y_t \ge \alpha^t  \text{ i.o.}\right) = 0$, showing that with probability $1$, $0\le  Y_t-\tilde Y_t \le\alpha^t $ eventually. Similar computations can be performed for an AR(p)-Poisson, showing that stating points are not important and that the convergence of
$Y_t-\tilde Y_t$ to $0$ is exponentially fast.

\subsection{Gaussian HMM}\label{app:hmm}
According to \cite{Nasri/Boucher/Perreault/Remillard/Huard/Nicault:2020} or \cite{Nasri/Remillard/Thioub:2024}, for a Gaussian Hidden Markov  model with $r$ regimes, the conditional distribution of the process $Y_t$ verifies $\disp G_t(y) = \sum_{j=1}^r \Phi\left(\frac{y-\mu_j}{\sigma_j}\right) Q_{\tau_{t-1},j}$, where $\Phi$ (resp. $\phi$) is the cdf (resp. pdf) of a standard Gaussian variable, where $\tau_t$ is the process of regimes with values in $\{1,\ldots,r\}$ and transition matrix $Q$. It is assumed that $\tau_t$ is stationary and ergodic, with stationary distribution $\pi$.  Hence, $G_t$ is stationary and $G_t$ converges to $\disp G_\infty(y)=\sum_{j=1}^r \pi_j \Phi\left(\frac{y-\mu_j}{\sigma_j}\right)  $.
It follows from the calculations in Remark \ref{rm:hyp} that $G_t^{-1}$ is also stationary. Next, for simplicity, take $r=2$, so the parameters are $\mu_1,\mu_2,\sigma_1,\sigma_2, Q_{11},Q_{21}$.
As a result, $\partial_{\mu_j} G_t(y) = -\frac{1}{\sigma_j}\phi\left(\frac{y-\mu_j}{\sigma_j}\right) Q_{\tau_{t-1},j}$, $\partial_{\sigma_j} G_t(y) = -\frac{1}{\sigma_j^2}\phi\left(\frac{y-\mu_j}{\sigma_j}\right) Q_{\tau_{t-1},j}$, $\partial_{Q_{11}} G_t(y) = \I\{\tau_{t-1}=1\}\left\{\Phi\left(\frac{y-\mu_1}{\sigma_1}\right)-\Phi\left(\frac{y-\mu_2}{\sigma_2}\right) \right\} $, and
$\partial_{Q_{21}} G_t(y) = \I\{\tau_{t-1}=2\}\left\{\Phi\left(\frac{y-\mu_1}{\sigma_1}\right)-\Phi\left(\frac{y-\mu_2}{\sigma_2}\right) \right\} $. It then follows that $\dot G_t$ is stationary. It follows that $\dot G_t(y)$ is Lipschitz in $Y$,  having  a bounded continuous density. Since $G_t^{-1}(u)\to G_\infty^{-1}(u)$, it follows that $
\left\| \bgamma_t(u) - \dot G_t \circ G_\infty^{-1}(u)\right\| \le c \left|G_t^{-1}(u)-G_\infty^{-1}(u)\right|$. By ergodicity, as $n\to\infty$,
$\disp
\frac{1}{n}\sum_{t=1}^n \left\| \bgamma_t(u) - \dot G_t \circ G_\infty^{-1}(u)\right\| \le  \frac{c}{n}\sum_{t=1}^n\left|G_t^{-1}(u)-G_\infty^{-1}(u)\right|\to 0$.